\documentclass[aip,jcp,preprint]{revtex4-1}
\usepackage[T1]{fontenc}
\usepackage[utf8]{inputenc}
\usepackage[english]{babel}
\usepackage{blindtext}
\usepackage{hyperref}
\usepackage[dvipsnames]{xcolor}
\usepackage{amsmath}
\usepackage{amssymb}
\usepackage{float}
\usepackage{afterpage}
\usepackage{subcaption}
\usepackage{graphicx}
\usepackage{tikz}
\usetikzlibrary{quotes, angles, positioning}
\usepackage{breqn}
\usepackage{ulem}

\DeclareGraphicsExtensions{.pdf}

\definecolor{newstuff}{rgb}{0,0,0}

\newcommand{\epsin} {\epsilon_{in}}
\newcommand{\epsout} {\epsilon_{out}}

\newcommand{\BETA}{ \frac{ \epsin }{ \epsout } }

\makeatletter
\let\cat@comma@active\@empty
\makeatother

\begin{document}

\title{A closed-form, analytical approximation for apparent
surface charge and electric field of molecules. }
\author{Dan Folescu}
\affiliation{Department of Computer Science, Virginia Tech}
\author{Alexey V. Onufriev}
\email{alexey@.cs.vt.edu}
\affiliation{Department of Computer Science, Virginia Tech}
\affiliation{Department of Physics, Virginia Tech}
\affiliation{Center for Soft Matter and Biological Physics, Virginia Tech, Blacksburg, VA 24061, USA}
\begin{abstract}
Closed-form, analytical approximations for 
electrostatic properties of molecules are of unique value, as these can
provide computational speed, versatility, and physical insight. Here, we 
derive a simple, closed-form formula for 
the apparent surface charge (ASC), as well as for the electric field, 
generated by a molecular charge distribution in aqueous solution. The
approximation, with no fitted parameters, is 
tested against numerical solutions of the
Poisson equation, where it yields a significant
speed-up. 
For neutral small molecules, the hydration free
energies estimated from the closed-form ASC formula 
are within $0.8$ kcal/mol RMSD from the
numerical Poisson reference; the electric field at the
surface is in quantitative agreement with the reference.
Performance of the approximation is also
tested on larger structures, including a protein, a DNA
fragment, and a viral receptor-target complex.
For all structures tested, a near quantitative agreement
with the
numerical Poisson reference is achieved, except in regions
of high negative curvature, where the new approximation is
still qualitatively correct. A unique efficiency feature of the
proposed ``source-based" closed-form approximation is that
the ASC and electric field can be estimated individually at any point or surface patch, 
without the need to obtain the full global solution.
An open source software implementation of the method is available:
\url{http://people.cs.vt.edu/~onufriev/CODES/aasc.zip}.
\end{abstract}
\maketitle

\section{Introduction}

Accurate and efficient modeling of solvation effects at the
atomistic level is a critical component of  modern efforts
to understand molecular structure and
function\cite{zhouElectrostaticInteractionsProteinCR2018,adcockMolecularDynamicsSurveyCR2006,karplusMolecularDynamicsSimulationsNSB2002,ramanSpatialAnalysisQuantificationJACS2015,amaroCommunityLetterRegardingJCIM2020}.
Analysis and visualization of electrostatic properties of biomolecules,
including the electric field and surface charge generated by the molecular
charge distribution, have made an impact in
qualitative reasoning about biomolecules
\cite{zhouElectrostaticInteractionsProteinCR2018,tolokhWhyDoublestrandedRNANAR2014}.

There are two broad approaches to the modeling of molecular electrostatics
and solvation effects: 
explicit and implicit solvation methods\cite{onufrievWaterModelsBiomolecularWIRCMS2017}.
Arguably the most widely used model of solvation is
that for which individual solvent molecules are treated explicitly,
on the same footing with the target molecule. However, accuracy of
the explicit solvent representation comes at high price,
computationally, limiting the practical utility of atomistic
simulations in many areas. 
The implicit, continuum solvation approach – treating solvent as a continuum with the dielectric and non-polar properties of water – can offer much greater effective simulation speeds compared to the explicit solvent models \cite{cramerImplicitSolvationModelsCR1999,honigClassicalElectrostaticsBiologyS1995,berozaCalculationsProtonbindingThermodynamicsME1998,maduraBiologicalApplicationsElectrostaticRiCC1994,gilsonTheoryElectrostaticInteractionsCOSB1995,scarsiContinuumElectrostaticEnergiesJPCA1997,luoAcceleratedPoissonBoltzmannJCC2002,simonsonElectrostaticsDynamicsProteinsRoPiP2003,bakerImplicitSolventElectrostaticsNAfMS2006,bardhanBiomolecularElectrostaticsWantCSD2012,liDielectricConstantProteinsJCTC2013}.
The Poisson equation\cite{honigClassicalElectrostaticsBiologyS1995,berozaCalculationsProtonbindingThermodynamicsME1998,rocchiaExtendingApplicabilityNonlinearJPCB2001,nichollsRapidFiniteDifferenceJCC1991,bakerElectrostaticsNanosystemsApplicationPNASUSA2001,chenMIBPBSoftwarePackageJCC2011,nguyenAccurateRobustReliableJCC2017}
of classical electrostatics\cite{jacksonClassicalElectrodynamics1999} provides an exact formalism -- within the continuum, local, linear-response dielectric approximation of solvent in the absence of mobile ions -- for computing the electrostatic potential $V(\mathbf{r})$ 
produced by a molecular charge distribution $\rho(\mathbf{r})$
characterizing the solute: 
\begin{align}\label{eq:Poisson}
\nabla \epsilon(\mathbf{r}) \nabla V(\mathbf{r}) = -4 \pi \rho(\mathbf{r}),
\end{align}
where $\epsilon(\mathbf{r})$ is the dielectric constant.
Once 
$V(\mathbf{r})$ is obtained, the electrostatic part of the solvation free energy
is easily computed\cite{jacksonClassicalElectrodynamics1999}.

The problem of finding $V(\mathbf{r})$ is mathematically  equivalent\cite{miertusElectrostaticInteractionSoluteCP1981,miertusApproximateEvaluationsElectrostaticCP1982}
to finding a continuous charge density, $\sigma$, on 
the dielectric boundary (DB), such that:
\begin{gather}\label{eq:standard_potential_two_dielectric}
  \begin{aligned} 
  V( \mathbf{r} ) &= \sum\limits_i \frac{ q_i }{ \vert \mathbf{r} - \mathbf{r}_i \vert } + \oint\limits_{\partial S} \frac{ \sigma( \mathbf{s} ) }{ \vert \mathbf{r} - \mathbf{s} \vert }\, d^2s,
  \end{aligned}
\end{gather}
where $\rho(\mathbf{r})$ is the discrete charge distribution,
formed by $n$ point charges $q_1, \cdots, q_n$, and $ \sigma(\mathbf{s})$ is the {\it apparent surface
charge} (ASC)  associated with each surface patch
$\mathbf{s}$. The second term in the above equation
represents the so-called reaction field
potential\cite{onufrievContinuumElectrostaticsSolventMSE2010,onsagerElectricMomentsMoleculesJACS1936,barkerMonteCarloStudiesMP1973}.
Conceptually, once the ASC, $\sigma(\mathbf{s})$, is found, all of the solvation
effects, at the level of the Poisson equation, can be computed. A great
variety of practical, widely used methods, including multiple modern 
derivatives of 
PCM\cite{miertusElectrostaticInteractionSoluteCP1981} and 
COSMO\cite{klamtConductorlikeScreeningModelJPC1995},  utilize this 
general idea -- the apparent surface charge (ASC) formalism, see Refs.
\cite{tomasiQuantumMechanicalContinuumCR2005,herbertDielectricContinuumMethodsWCMS2021} for comprehensive reviews. 

The reformulation of the Poisson problem via equation
\ref{eq:standard_potential_two_dielectric} has a number of
technical advantages made apparent over the years, especially in quantum
mechanical (QM) calculations\cite{cramerImplicitSolvationModelsCR1999,herbertDielectricContinuumMethodsWCMS2021}.

A number of ASC-based methods yield numerically exact solutions to the
Poisson equation, in the sense that the exact solution can, at least in
principle, be approximated with an arbitrary precision. Formally exact, linear-scaling implementations of
numerical ASC methods, based on conjugate
gradient or domain decomposition,
exist\cite{herbertDielectricContinuumMethodsWCMS2021}. Concerns related 
to computational 
cost of numerically exact 
approaches\cite{tomasiQuantumMechanicalContinuumCR2005} have led to the development of 
approximate ASC-based
methods, such as the widely
used COSMO \cite{klamtConductorlikeScreeningModelJPC1995},
GCOSMO \cite{truongNewMethodIncorporatingCPL1995}, and
C-PCM \cite{cossiAnalyticalSecondDerivativesJCP1998}. These
methods rely on approximations to equation
\ref{eq:standard_potential_two_dielectric}.  
Still,
even these approximate ASC-based methods employ a significant numerical
component, such as numerical matrix inversion, which may 
carry appreciable computational overhead, especially as the structure size
grows\cite{klamtCOSMONewApproachJCSPT21993,gordonAnalyticalApproachComputingJCP2008}.
Therefore, 
in applications where computational efficiency and algorithmic simplicity
is paramount,  numerical ASC-based methods 
may  not be as competitive
as approximations to the Poisson equation 
based purely on
closed-form analytical
expressions\cite{onufrievContinuumElectrostaticsSolventMSE2010}.

Among the fully analytical approximations to the
Poisson equation, the generalized Born (GB)
model\cite{hoijtink1956reduction,tucker1989generalized,cramerImplicitSolvationModelsCR1999,Still1997,Jayaram1998,Still1990,Onufriev2000,Dominy1999,Bashford2000,Calimet01,Hawkins95,Hawkins1996,Schaefer1996,Feig_Im04,MLee2002,Lee0803,Onufriev0504,Wang03,Gallicchio04,Nymeyer03,Ghosh98,Scarsi97,Im03,Haberthur08,Grant07,Zhou07,Labute08,Onufriev1102,Onufriev2011,Lange2012}
is
arguably the most widely used, especially in atomistic
simulations\cite{Simmerling2002,Chen06,Lei07,Pitera03,Jagielska07,Hornak:2006:Proc-Natl-Acad-Sci-U-S-A:16418268,Amaro09,JZmuda06,Chocholousova06,Zheng04,ZhuAlexov2005}.  
However, despite its multiple documented success
stories\cite{onufrievGeneralizedBornImplicitARB2019} the
GB model does not have the versatility of 
equation \ref{eq:standard_potential_two_dielectric}, and the
associated benefits of an
ASC-based formulation of biomolecular electrostatics.
	Here we aim to fill the gap by deriving an {\it analytical}, closed
form approximation to the Poisson equation for the ASC and the (normal)
electric field around an arbitrary shaped molecule. Standard numerical
solutions of the Poisson equation are used as the reference. 
We refrain from comparisons with  well-established,
optimized numerical implementations of ASC methods in this initial investigation. 


The outline of the paper is as follows.
Section \ref{sec:computational_details} describes computational testing materials and methodology.
Section \ref{sec:theory_and_results} is focused on our analytical ASC
approximation: we first derive an exact analytical ASC reference on a sphere (\ref{ssec:EPB}),
and present our approximate form of the ASC for arbitrary molecular geometries (\ref{ssec:AASC}).
We test the approximate ASC against the exact analytical ASC reference using a spherical test case (\ref{ssec:apb_test}), which simulates relevant electrostatic configurations that we will encounter in later sections.
Application of our model to solvation energy calculations is presented in section \ref{ssec:ESFE_ASC}.
Numerical performance and accuracy  analysis, along with an example 
applications are presented in section \ref{sec:numerical_applications_results}, where we first examine the computational speed of our method (\ref{ssec:computational_speed}, before testing its accuracy on small and large molecules (\ref{sssec:sm_test}, \ref{sssec:lm_test}).
Finally, we showcase our analytical ASC approximation on a presently relevant biomolecular complex: that of the human ACE2 receptor and SARS-CoV-2 spike glycoprotein (\ref{ssec:6M0J}).

\section{Methods}\label{sec:computational_details}

\subsection{Structures}

For testing, we utilize a set of 173 neutral small molecules from
version 0.52 of FreeSolv
database\cite{mobleyFreeSolvDatabaseExperimentalJCAMD2014,mobleyCorrectionSmallMoleculeJCTC2015}.
The original set of nearly 600 molecules 
was narrowed down to include only those molecules containing
hydrogen, oxygen, nitrogen, and carbon atoms.
The small molecules under consideration are all rigid
- having small conformational variability as seen in
  molecular dynamic (MD) simulations
\cite{mukhopadhyayIntroducingChargeHydrationJCTC2014}. The
choice of rigid molecules allows us to focus on the physics
of solvation, while mitigating the uncertainty related to
conformational sampling.
\textit{ambpdb} \cite{caseAmberBiomolecularSimulationJCC2005} was used to generate PQR format files from AMBER format coordinate and topology files \cite{mobleyFreeSolvDatabaseExperimentalJCAMD2014}.
Additionally, two larger biomolecules are used: a 25 bp
poly-A B'-form
dsDNA\cite{tolokhWhyDoublestrandedRNANAR2014}; and the hen-egg lysozyme (PDB:2LZT) \cite{ramanadhamRefinementTriclinicLysozymeACB1990}.
We also test  our method on a portion of the ACE2/SARS-CoV-2 complex (PDB:6M0J) \cite{wangCrystalStructureSARSCoV22020} receptor binding domain (RBD).
6M0J RBD residues were determined through A/E chain contacts within 3.8 \AA.
In each chain, residues within 1.5 \AA~of the contacts were also included.
H++ server \cite{gordonServerEstimatingPKasNAR2005} was used to generate 
protonated PQR format files.
PQR format files for small molecules and the SARS-CoV-2 complex, along with RBD contact residue lists, are provided in the accompanying code package, see the abstract.

\subsection{Dielectric Boundary Representations}\label{ssec:DB}
Just like the numerical Poisson solvers, ASC-based
methods rely on  molecular boundary representations
\cite{connollyAnalyticalMolecularSurfaceJAC1983,decherchiGeneralRobustRayCastingBasedPO2013,grantSmoothPermittivityFunctionJCC2001}. 
These representations of the actual
molecular shape are crucial to the accuracy and, to some extent, 
efficiency of modern
implicit solvation models.  
The question of which definition of the dielectric boundary (DB) is most 
appropriate, is non-trivial; there is no universal 
default\cite{tjongDielectricBoundaryPoissonBoltzmannJCTC2008,cramerUniversalApproachSolvationACR2008,onufrievAccuracyContinuumElectrostaticJTCC2014}.

Here, our main goal is to assess the accuracy of 
the new ASC approximation against its primary accuracy metric, the 
numerical PB. Both methods can utilize whichever DB one considers most
appropriate for the given application. Our secondary goal is to compare the
new ASC approximation to a fast analytical GB model. These goals dictate
the choice of the DB representation chosen here.

Within our ASC method we 
approximate the solute-solvent interface -- the
DB --  using the solvent
excluded surface (SES)
\cite{richardsAreasVolumesPackingARBB1977,connollyAnalyticalMolecularSurfaceJAC1983}, with Bondi
\cite{bondiVanWaalsVolumesJPC1964} atomic radii and
a water probe of $1.4$ \AA.
This DB is triangulated with the open-source package, NanoShaper \cite{decherchiGeneralRobustRayCastingBasedPO2013}.
In each relevant test, we match the NanoShaper grid spacing with
that used by the numerical PB reference; we use $0.1$
\AA~for small molecules accuracy comparisons, 
$0.25$ \AA~for fair speed comparisons, and $0.5$ \AA~for large molecule
electric field normal comparisons.
The two existing reference methods employed in this work, the NBP and GB,
sections \ref{ssec:NPBR} and \ref{ssec:GBR}, use matching dielectric
boundaries based on the same set of Bondi radii with 1.4 \AA~water probe
radius (unless otherwise specified), for consistency.

\subsection{Numerical Poisson-Boltzmann Reference}\label{ssec:NPBR}

\begin{figure}[H]
    \centering
    \resizebox{0.75\linewidth}{!}{
      \includegraphics{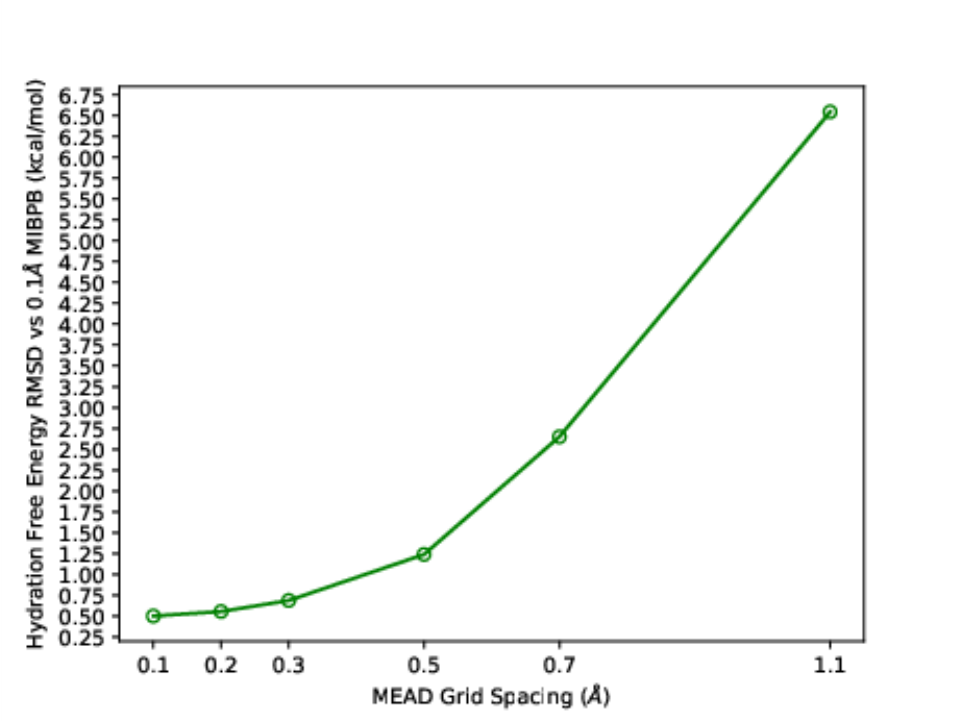}
    }
    \caption{
      Convergence of hydration free energies computed by MEAD. The
convergence is assessed against reference values computed by
MIBPB\cite{chenMIBPBSoftwarePackageJCC2011}, as RMSD over our small molecule set.
      MIBPB hydration free energies are computed with a 1.4\AA~water probe radius at  0.1 \AA~grid spacing.
      MEAD hydration free energies are computed with identical parameters, except for the variable grid spacing. 
      MIBPB small molecule electrostatic hydration free energies ranged from -0.05 to -14.37 kcal/mol, with an average magnitude of 6.68 kcal/mol.
    }
    \label{fig:mead_mibpb_convergence}
  \end{figure}

For numerical Poisson-Boltzmann (NPB) reference calculations, we use the Macroscopic Electrostatics with Atomic Detail (MEAD) package \cite{bashfordObjectorientedProgrammingSuiteSCOPE1997}.
MEAD is a volumetric, finite-difference solver that can compute potential
maps and hydration energies, utilizing an SES DB representation.
The package was chosen primarily  
because it interfaces well with the visualization and analysis 
utility GEM\cite{gordonAnalyticalApproachComputingJCP2008} 
used, here to process the NPB-generated potential maps.
Using GEM, visualization of the electric field around the solute can be achieved at any given distance from the DB.
We have verified convergence\cite{nguyenAccurateRobustReliableJCC2017} 
of MEAD-generated hydration free energies, Figure
\ref{fig:mead_mibpb_convergence}, and found it acceptable for our purposes 
at fine enough grid resolutions of interest to us in this work. A modern, 
2nd order method MIBPB\cite{zhouHighlyAccurateBiomolecularJCC2008} was used
as the accuracy reference for MEAD.

Based in part on  the convergence analysis, we perform potential-map calculations at a grid spacing of $0.1$ \AA, and hydration energy calculations at an inner and outer bounding-box grid spacing of $0.1$ \AA~and $0.5$ \AA, respectively, for small molecule accuracy comparisons.
NPB reference solvation energy calculations utilize inner and outer dielectrics of $1$ and $80$, respectively.
The water probe radius is 1.4\AA, unless otherwise stated.
For fair speed comparisons, we perform potential-map calculations at a grid spacing of $0.25$ \AA, and hydration energy calculations at a single bounding-box grid spacing of $0.25$ \AA, considered standard for finite-difference NPB calculations.
For larger molecules, we perform potential-map calculations at a grid spacing of $0.5$ \AA, and hydration energy calculations at an inner and outer bounding-box grid spacing of $0.5$ \AA~and $1.0$ \AA, respectively.
The total number of NPB grid points was determined by setting a volumetric bounding-box side length slightly larger (+1 \AA) than the maximal intra-molecular distance.
We numerically approximate the electric field normal at a point $\mathbf{r}$ by a two-point stencil:
\begin{align}
  \mathbf{E}_\perp(\mathbf{r}) = - \frac{\partial V}{\partial \vec{n}} \approx \frac{ V (\mathbf{r} + h \hat{n}) - V (\mathbf{r} - h \hat{n}) }{2h},
\end{align}
where $\mathbf{r} + h \hat{n}$ and $\mathbf{r} - h \hat{n}$
are two sampling points distance $r \pm h$ from the DB along the surface normal $\hat{n}$;~
see Ref. \cite{gordonAnalyticalApproachComputingJCP2008} for
additional details of the sampling protocol.
Here, $h$ was chosen to minimize the distance between sampled  points,
while still being large enough so that the sampled points are
distinct. Notice that if $h$ were too small, the sampling
protocol would sample the exact same two points of the cubic lattice used
in the NPB reference calculations. The largest possible distance between two such
grid points is the diagonal of the cubic grid, which determines the minimum
$h$.
To illustrate this numerical constraint, $h$ must be larger than $\sqrt{3} / (0.1)^2 \sim 0.173$ \AA~ for members of the small molecule dataset, whose potential grids were computed with the NPB reference at a grid spacing of $0.1$ \AA.

Additionally, to avoid numerical artifacts of the NPB reference near the
DB, and mitigate  possible effects of minor differences between 
the internal representations of the SES computed by our NPB reference 
and NanoShaper, the field is
computed a distance $p > h$ from the DB, Figure \ref{fig:derivationSetup}:
doing so ensures that we do not accidentally sample grid points inside the
molecule.
The need to use a non-zero \textit{projection distance} in
the NPB calculations makes it necessary to
consider the electric field normal values near the DB as the numerical
reference for assessing the accuracy of the analytical ASC, 
rather than the apparent surface charge itself (which, up
to a prefactor, is essentially the normal
component of the field right at the DB, see equation \ref{eq:sigma}).

\subsection{Generalized-Born Solvation Free Energy Reference}\label{ssec:GBR}

We utilize the
IGB5\cite{onufrievExploringProteinNativePSFB2004} GB model
from AMBER package\cite{caseAmberBiomolecularSimulationJCC2005}.
This model was parameterized against reference NPB hydration free energies calculated based on SES surface, Bondi radii, and 1.4 \AA~water probe radius.

\subsection{Accuracy Metrics Used}

We test our ASC approximation, first against the exact PB (EPB, see section \ref{ssec:EPB} ) reference, and then against NPB and GB references.
Our comparison with the EPB reference is used directly for the apparent
surface charge, employing the two test charge configuration in Figure
\ref{sfig:two_charge_test}.
Per-vertex electric field normal values are compared against the NPB reference, averaging over each vertex in a given biomolecule, and over each biomolecule in the comparison set.
Electrostatic hydration free energies are compared against the NPB and GB
references with inner and outer dielectrics $\epsilon_{in} = 1$ and $\epsilon_{out} = 80$, unless otherwise specified.
All results will be in $e/$\AA$^2$ for apparent surface charge, kcal/$( \text{ mol } \cdot e \cdot \text{\AA})$ for electric field normals, and kcal/mol for electrostatic hydration free energies.

\subsection{Computer Specifications}

All computations and visualizations were completed on a commodity desktop computer with an Intel Core i7 (or equivalent) processor, using a maximum of 32 GB of memory.

\section{Theory and Results}\label{sec:theory_and_results}

\subsection{Preliminaries: Analytical Poisson-Boltzmann Reference}\label{ssec:EPB}

\begin{figure}[H]
		\resizebox{\linewidth}{!}{
      \begin{subfigure}{0.5\textwidth}
        \subcaption{}
        \includegraphics[width=\textwidth]{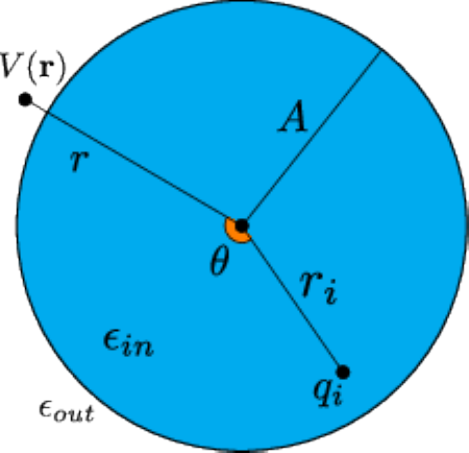}
        \label{sfig:one_charge_description}
      \end{subfigure}
      \quad\quad
      \begin{subfigure}{0.5\textwidth}
        \subcaption{}
        \includegraphics[width=\textwidth]{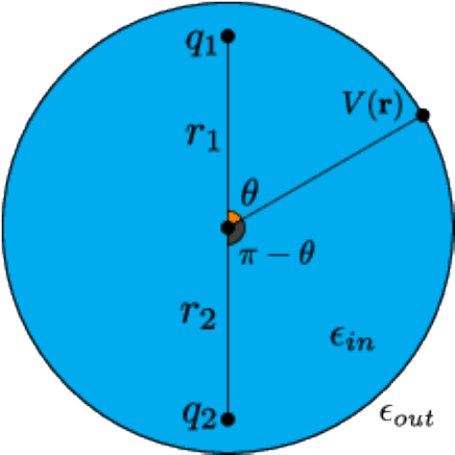}
        \label{sfig:two_charge_test}
      \end{subfigure}
		}
	\caption{
    Geometry and charge settings for the Kirkwood
    multipolar expansion, \ref{sfig:one_charge_description},
    and for the two-charge test case, \ref{sfig:two_charge_test}.
    These notations and geometries are used in equations \ref{eq:KirkwoodInfinite} and \ref{eq:KirkwoodInfiniteNormalDerivative}.
    In each case, a perfectly spherical, sharp DB is utilized, with $\epsilon_{in}$ and $\epsilon_{out}$ denoting inner and outer dielectric constants, respectively.
    The angle $\theta$ (resp. $\pi - \theta$) is subtended by the lines connecting the point of observation, $\mathbf{r}$, and the charge(s) to the spherical center.
    In panel \ref{sfig:one_charge_description}, a single point charge, $q_i$, is located $r_i$ away from the center of a spherical boundary with radius $A$.
    In panel \ref{sfig:two_charge_test}, two point charges, $q_1$ and $q_2$, are located on the vertical diameter of the sphere.
    The charges are of equal distance, $r_1 = r_2$, from the spherical center.
    The electrostatic potential $V(\mathbf{r})$ is computed at the point of observation $\mathbf{r}$.
	}
	\label{fig:KW_description_test}
\end{figure}

In the context of implicit solvation, the
simplest scenario is that of a solute with a sharp,
spherical DB, Figure \ref{sfig:one_charge_description}.
\noindent
For such a spherical boundary, as in Figure
\ref{sfig:one_charge_description}, Kirkwood
\cite{kirkwoodTheorySolutionsMoleculesJCP1934} gave the
exact, analytical solution of equation \ref{eq:Poisson} for
the potential $V_i$ at the DB due to a single charge $q_i$ inside the
boundary. Here, we use the solution valid on or exterior to the
spherical DB\cite{fenleyAnalyticalApproachComputingJCP2008}, 
without consideration for mobile ions. At $r = A$: 
\begin{align} \label{eq:KirkwoodInfinite}
V_i &= - \frac{q_i}{A} \left( \frac{1}{\epsilon_{in}} - \frac{1}{\epsilon_{out}} \right) \sum\limits_{l=0}^{\infty} \left[ \frac{1}{1 + \left( \frac{l}{l+1} \right) \left( \BETA \right) } \right] \left( \frac{r_i}{A} \right)^l P_l(\cos \theta) + \frac{q_i}{A} \left( \frac{1}{\epsilon_{in}} \right) \sum\limits_{l=0}^{\infty} \left( \frac{r_i}{A} \right)^l P_l(\cos \theta).
\end{align}
The apparent surface charge $\sigma$ is related to the normal component of the electric field  $\mathbf{E}_{\perp} =  \left( \frac{ \partial V }{ \partial \vec{n} } \right)_{out}$ at (just outside) the boundary via:\cite{miertusElectrostaticInteractionSoluteCP1981,jacksonClassicalElectrodynamics1999}
\begin{align}\label{eq:sigma}
\sigma &= \frac{ 1 }{ 4 \pi } \left( \frac{ \epsilon_{out} }{ \epsilon_{in} } - 1 \right) \left( \frac{ \partial V }{ \partial \vec{n} } \right)_{out}.
\end{align}
From this, and equation \ref{eq:KirkwoodInfinite}, we obtain
an {\it exact}, analytical expression for the apparent
surface charge on the spherical DB:
\begin{dgroup} \label{eq:KirkwoodInfiniteNormalDerivative}
\begin{dmath*}
\sigma_{KW} = \sum\limits_i \frac{q_i}{4 \pi} \left( \frac{1}{\epsilon_{in}} - \frac{1}{\epsilon_{out}} \right) \left[ \sum\limits_{l=0}^{\infty} \left[ \frac{1}{1 + \left( \frac{l}{l+1} \right) \left( \BETA \right) } \right] ( l + 1 ) \left( \frac{ r_i^l }{ A^{ l + 2 } } \right) P_l(\cos \theta) - \left( \frac{ 1 }{\epsilon_{in}} \right) \sum\limits_{l=0}^{\infty} ( l + 1 ) \left( \frac{ r_i^l }{ A^{ l + 2 } } \right) P_l(\cos \theta) \right],
\end{dmath*}
\end{dgroup}
where the summation is over all of the enclosed charges. 
Equation \ref{eq:KirkwoodInfiniteNormalDerivative} will
provide a key check for our analytical ASC approximation.

\subsubsection{Convergence Analysis of $\sigma_{KW}$}\label{sssec:EPB_convergence}

The presence of the indexing term, $( l + 1 )$, is notable
in its effects on the convergence characteristics of
equation \ref{eq:KirkwoodInfiniteNormalDerivative}, by
increasing the number of terms necessary to obtain
a converged, accurate reference.
As the ratio $ r_i/A $ approaches $1$ - that is, as the
charge approaches the DB - slow convergence of the
approximate solution manifests itself.
\begin{figure}[H]
	\begin{center}
		\resizebox{\linewidth}{!}{
		\includegraphics[width=\linewidth]{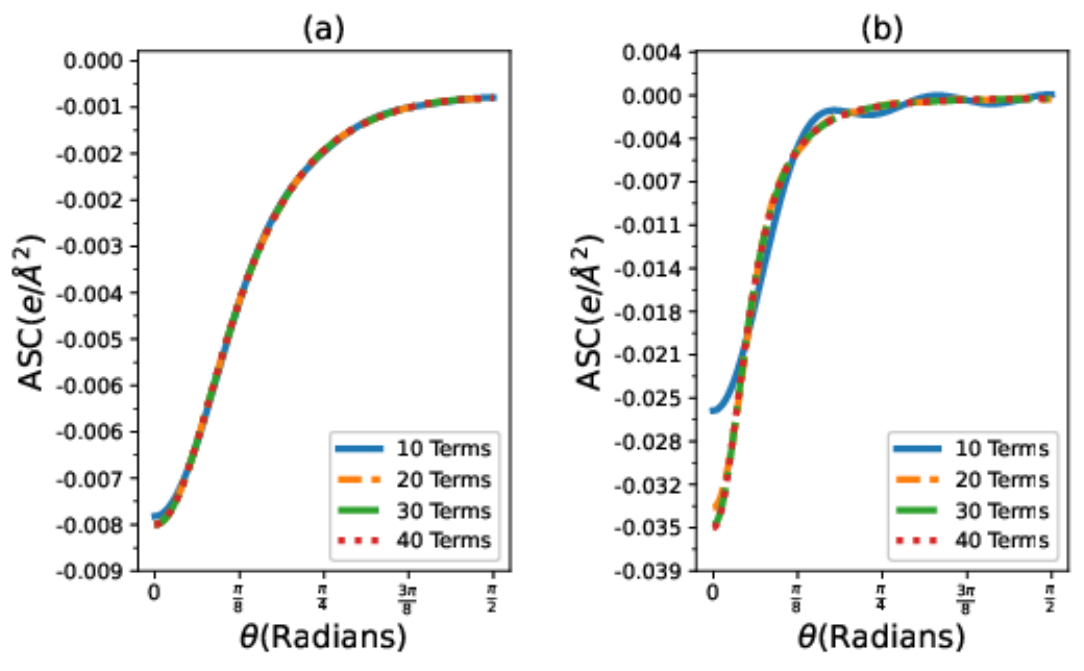}
		}
	\end{center}
	\caption
	{
The apparent surface charge (ASC, per unit area) computed using the
truncated, infinite series  analytical solution of the
Poisson problem on a sphere, 
equation \ref{eq:KirkwoodInfiniteNormalDerivative}.
The convergence tests are conducted on the dual-positive
test case, Figure \ref{sfig:two_charge_test}.
The ASC is sampled at the spherical boundary $A = 10$ \AA~away from the center an angle $\theta$ from $q_1$.
Panel (a) $r_1 = r_2 = 6$ \AA; (b) $r_1 = r_2 = 8$ \AA; see Figure \ref{sfig:two_charge_test}.
Partial sums of the infinite series solution, equation \ref{eq:KirkwoodInfiniteNormalDerivative}, with $M = 10,20,30,40$ terms are examined.
The truncated sums are shown with blue, orange, green, and red lines, respectively.
	}
	\label{fig:kirkwoodExactTest}
\end{figure}
In Panel (b) of Figure \ref{fig:kirkwoodExactTest}, we see that, even for $\frac{r_i}{A} = 0.8$, it is possible to achieve both qualitatively and quantitatively reasonable results.
The analytical reference appears well-converged, Figure \ref{fig:kirkwoodExactTest}, for our purposes, at $M=30$ terms.
Hence, we numerically approximate equation \ref{eq:KirkwoodInfiniteNormalDerivative} by truncating to the first $M = 30$ terms, calling the resulting expression the \textit{essentially exact Poisson-Boltzmann (EPB) reference}, which we use further in this work.

\subsection{Main Result: Analytical Apparent Surface Charge}\label{ssec:AASC}

\begin{figure}[H]
	\begin{center}
		\resizebox{0.75\linewidth}{!}{
		\includegraphics{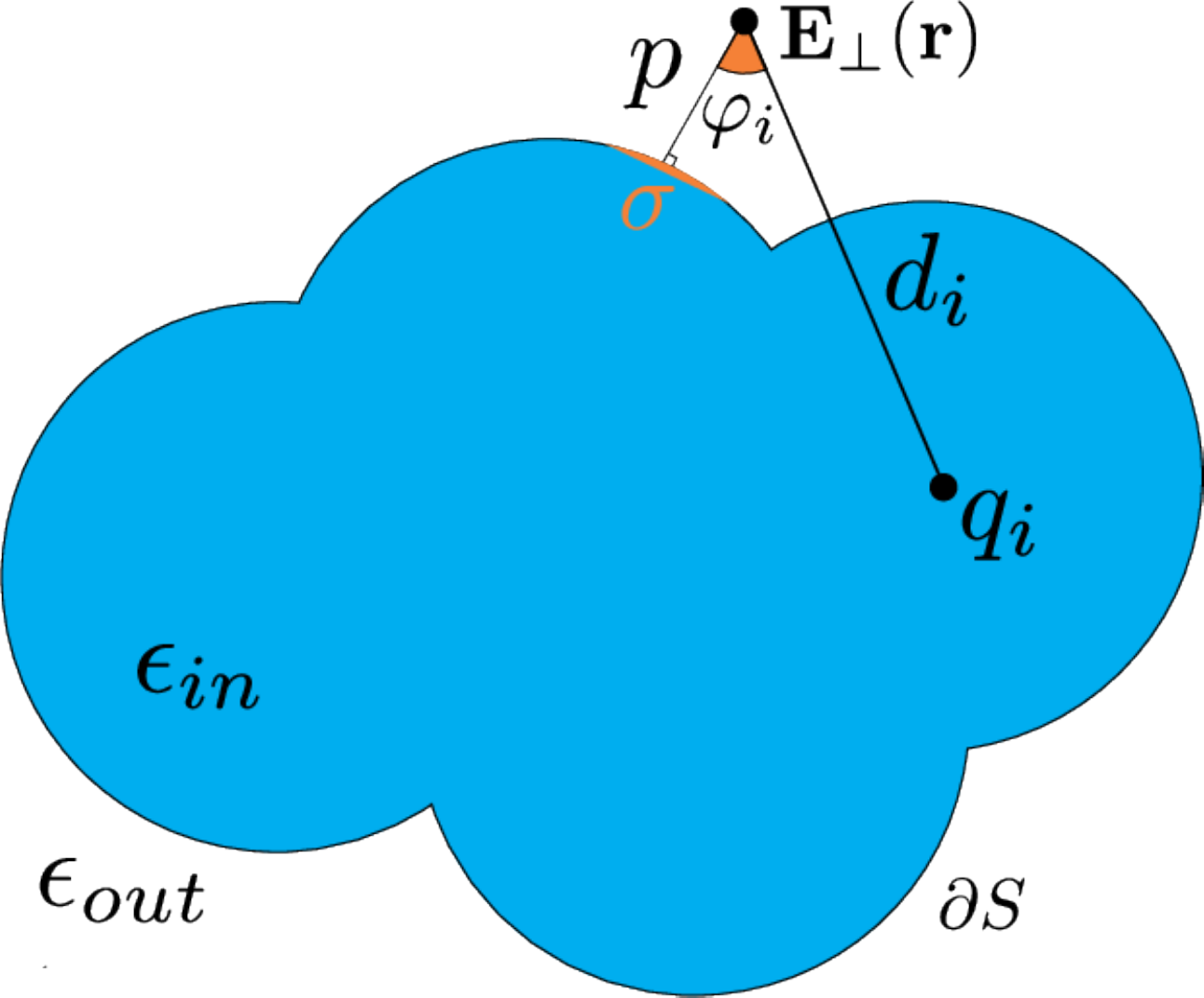}
		}
	\end{center}
	\caption{
The two dielectric problem for an arbitrary molecule $S$ with smooth boundary $\partial S$.
The boundary separates the inner (blue) and outer (white) dielectric regions, with constants $\epsilon_{in}$ and $\epsilon_{out}$, respectively. 
For a non-spherical DB, the distance $r$ from the spherical center, Figure
\ref{sfig:one_charge_description}, to the sampling point $\mathbf{r}$ is 
replaced with the generalized expression $r = A+p$.
Here, $A$ is the so-called \textit{electrostatic size} of $S$, which
characterizes the dimension of the molecule~\cite{sigalovAnalyticalElectrostaticsBiomoleculesJCP2006}, and $p$ denotes the \textit{projection distance} from $\mathbf{r}$ to $\partial S$ along the surface normal, $\vec{n}$.
$q_i$ is the source charge under consideration,
with $d_i$ denoting the distance between 
$q_i$ and $\mathbf{r}$.
When $p > 0$, the \textit{electric field normal} $\mathbf{E}_{\perp}( \mathbf{r} )$ of $S$ at $\mathbf{r}$ is computed via equation \ref{eq:electric_field_normal}.
When $p = 0$, the ASC, $\sigma(\mathbf{r})$, is computed via equation \ref{eq:sigma_total}.
	}
	\label{fig:derivationSetup}
\end{figure}

To derive our ASC approximation, we begin with the
previously derived closed-form approximation for the electrostatic potential
around (outside) an arbitrary molecular
shape\cite{fenleyAnalyticalApproachComputingJCP2008}, see
Figure \ref{fig:derivationSetup}:
\begin{align}\label{eq:fenley_potential}
\begin{aligned}
	V_i \approx \left( \frac{ q_i }{ \epsilon_{out} \left( 1 + \alpha \left( \BETA \right) \right) } \right) \left[ \frac{(1 + \alpha)}{d_i} - \frac{\alpha \left( 1 - \BETA \right) }{ r } \right].
\end{aligned}
\end{align}
We utilize the polar orthonormal frame, $e_r = \frac{ \partial }{ \partial r }\,;\,e_\theta = \frac{ 1 }{ r } \frac{ \partial }{ \partial \theta }$, to take its derivative, for use in equation \ref{eq:sigma}.
The derivative vanishes in the direction of $e_\theta$,
yielding:
\begin{align}\label{eq:point_charge_potential_normal_derivative}
  \mathbf{E}_{\perp}(\mathbf{r}) = - \frac{ \partial V_i }{ \partial \vec{n} } &= - \left[ \frac{ \partial V_i }{ \partial r } - \cos( \varphi_i ) \frac{ \partial V_i }{ \partial d_i } \right].
\end{align}
Exploiting the geometry in Figure \ref{fig:derivationSetup}, we relate $\cos( \varphi_i )$ as a dot product of the surface unit normal, $\hat{n}$, and the vector from $\mathbf{E}_{\perp}(\mathbf{r})$ to $q_i$, which we denote $\vec{ d_i }$ : $\cos( \varphi_i ) = \left( \hat{n} \cdot \vec{ d_i } \right) / d_i $.
Applying equation \ref{eq:fenley_potential} to
\ref{eq:point_charge_potential_normal_derivative}, and
summing over the charge distribution (Figure
\ref{fig:derivationSetup}) we arrive at:
\begin{align}\label{eq:electric_field_normal}
\begin{aligned}
  \mathbf{E}_{\perp}(\mathbf{r}) &= - \left( \frac{ 1 }{
\epsilon_{out} \left( 1 + \alpha \left( \BETA \right) \right) } \right)
\sum\limits_i q_i \left[ \left( \frac{ \alpha \left( 1 - \BETA \right) }{
(A+p)^2 } \right) - \cos( \varphi_i ) \left( \frac{(1 + \alpha) }{ d_i^2 }
\right)  \right],
\end{aligned}
\end{align}
where we have made the substitution $r = A+p$ described in Figure \ref{fig:derivationSetup}.
At the dielectric boundary, $p=0$, applying equation
\ref{eq:sigma} to equation \ref{eq:electric_field_normal}
gives the following closed-form, analytical approximation
for the apparent surface charge:
\begin{align}\label{eq:sigma_total}
\begin{aligned}
  \sigma = - \left( \frac{ 1 }{ \epsilon_{in}
} - \frac{ 1 }{ \epsout } \right) \left( \frac{ 1 }{ 4 \pi
\left( 1 + \alpha \left( \BETA \right) \right) } \right)
\sum\limits_i q_i \left[ \left( \frac{ \alpha \left(
1 - \BETA \right) }{ A^2 } \right) - \cos( \varphi_i ) \left( \frac{ (1 + \alpha) }{ d_i^2 } \right) \right].
\end{aligned}
\end{align}
These two equations are the main analytical result of this work. 
Equations \ref{eq:fenley_potential}, \ref{eq:electric_field_normal}, and \ref{eq:sigma_total} rely on approximating the $\frac{l}{l + 1}$ term
of equation \ref{eq:KirkwoodInfinite} as $\frac{l}{l + 1} = const  = \alpha$ for $l > 0$, 
which allows the infinite Kirkwood series to
be summed without truncation.
That approach proves critical to both accuracy and computational efficiency\cite{fenleyAnalyticalApproachComputingJCP2008} of the closed-form approximations, based on the Kirkwood solution.
The value of $\alpha$ in equation \ref{eq:fenley_potential} was rigorously
derived previously\cite{sigalovIncorporatingVariableDielectricJCP2005} to minimize RMSD to the exact Kirkwood solution for electrostatic potential, equation \ref{eq:KirkwoodInfinite}, assuming a random charge distribution inside a perfect sphere.
Here we have attempted to re-optimize the value of $\alpha$ in 
the context of equation \ref{eq:sigma_total}, aiming at best 
agreement with the reference for hydration free energies of our
small molecule set.
The effort led to only a very minor improvement in accuracy (not shown), so we have decided to retain the original\cite{sigalovIncorporatingVariableDielectricJCP2005} $\alpha = 0.580127$ for use in our ASC approximation.

\subsubsection{A Self-Consistency Check}\label{ssec:sanity_check}

\begin{figure}[H]
	\begin{center}
	\resizebox{\linewidth}{!}{
	\includegraphics{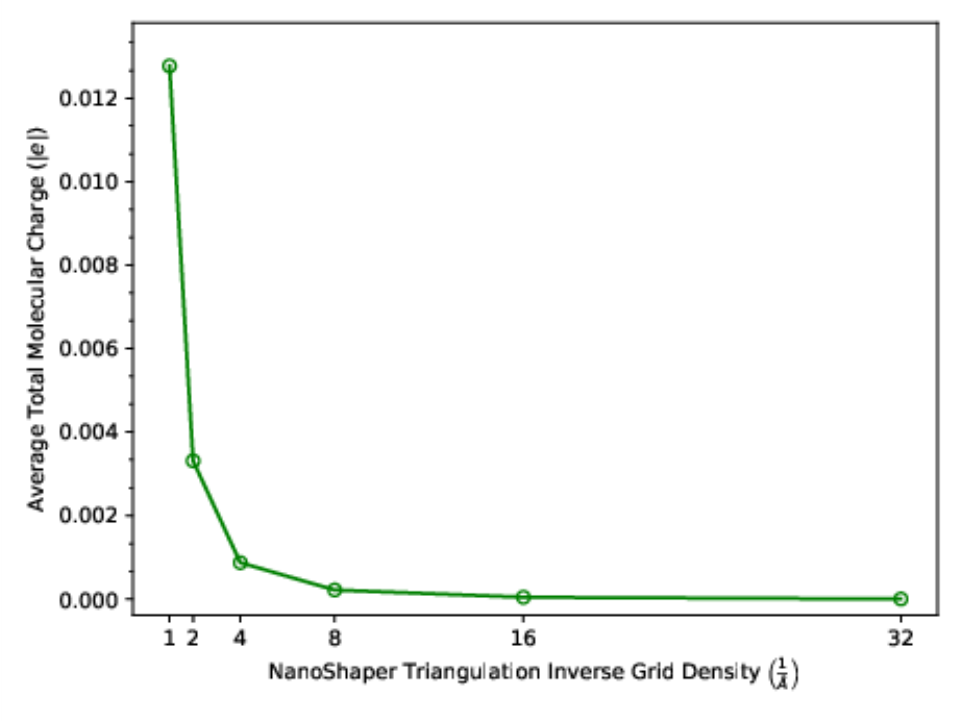}
	}
	\end{center}
	\caption
	{
Self-consistency and surface triangulation convergence 
check of our ASC method. The exact result for the total surface charge is
zero; as NanoShaper grid density is increased, the molecular
charge estimated via our implementation  also tends to zero. An average
over the entire set of small neutral molecules is shown.
NanoShaper inverse grid spacings are given in \AA$^{-1}$, while average total molecular charges are given in $ \vert e \vert $.
	}
	\label{fig:sanityCheck}
\end{figure}

Arguably the simplest self-consistency check of analytical ASC is that the
total surface charge produced by equation \ref{eq:sigma_total} should be
zero for any of the neutral small molecules making up our main test set:
\begin{align}\label{eq:discrete_infiniteDielectric_gauss}
\oint_{\partial S} \sigma\,d^2 s = 0.
\end{align}
In the discrete DB case, as the triangulation density is increased, we expect
the numerical approximation of the total charge integral, equation
\ref{eq:discrete_infiniteDielectric_gauss}, to approach zero.
Thus, the same simple check automatically tests both the analytical ASC and the DB discretization (triangulation) used.
As seen from Figure \ref{fig:sanityCheck}, our ASC implementation follows
the expected trend.  We stress that the only purpose of this simple test is a
``sanity check" of our code implementation; the accuracy of the derived
approximation is tested thoroughly below.  


\subsection{Accuracy against the Analytical PB Reference}\label{ssec:apb_test}

\begin{figure}[H]
	\begin{center}
	\resizebox{\linewidth}{!}{
	\includegraphics{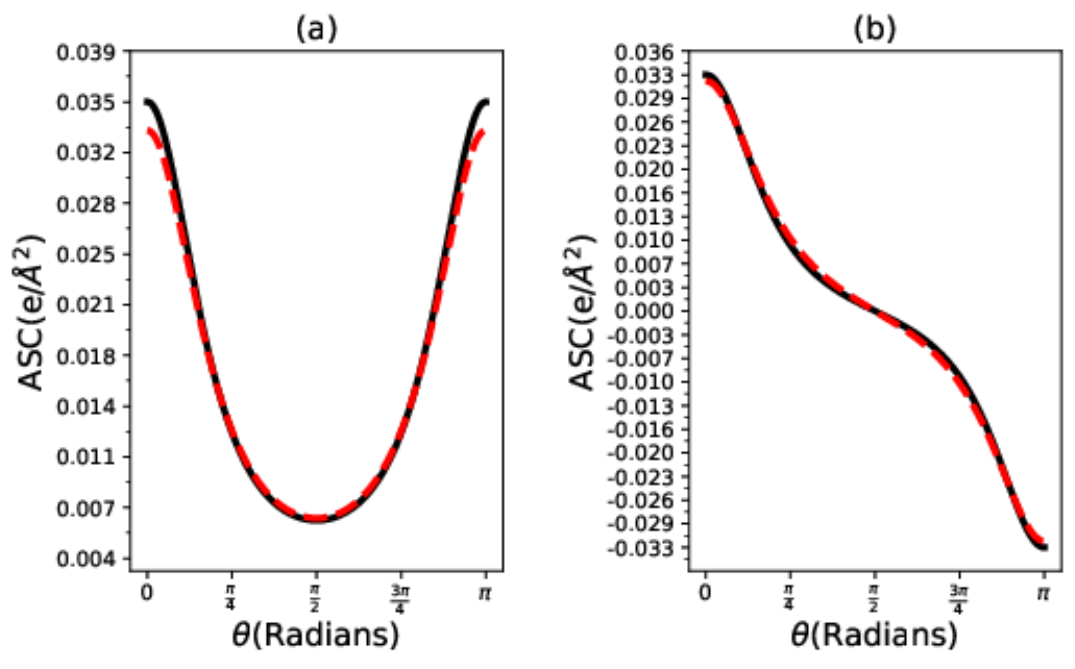}
	}
	\end{center}
	\caption
	{
Apparent surface charge of our ASC approximation (dashed red) and the EPB
reference (solid black) on the two-charge test distribution, shown in
Figure \ref{sfig:two_charge_test}. 
Panel (a): surface charge due to two point charges, $1.5$ \AA~away from the
boundary of a dielectric sphere of radius 3 \AA, with $q_1 = q_2 = 
-0.65$.
These specific parameters are intended to mimic two
oxygen atoms in a small molecule, with respect to the distance between a 
highly charged ``surface" atom and the DB.
Panel (b): the same charge distribution as in panel (a), but with $q_1 =
-0.65, q_2 = + 0.65$. 
Points were sampled from $0$ to $\pi$ in 0.0001 radian steps.
	}
	\label{fig:DipoleTest}
\end{figure}
From Figure \ref{fig:DipoleTest}, our approximation matches
the essentially exact Kirkwood solution, the truncation of equation \ref{eq:KirkwoodInfiniteNormalDerivative}, quite well on the two-charge test distribution, with only a slight, and expected, drop in accuracy closest to the point charges.
For $\theta \in [0, \pi]$, an RMSD of $0.00124 e/$\AA$^2$ and $0.00117 e/$\AA$^2$ was achieved between our ASC
approximation and the EPB reference, on the geometries described in panels (a) and (b) of Figure \ref{fig:DipoleTest}, respectively. 

\subsection{Electrostatic Solvation Free Energy with ASC}\label{ssec:ESFE_ASC}

The apparent surface charge formulation allows us to gain insights into a
variety of solvation effects \cite{onsagerElectricMomentsMoleculesJACS1936,
cammiRemarksUseApparentJCC1995}, including the estimate of 
\textit{hydration free energy}, which we will use extensively here to
evaluate the accuracy of our new approach against accepted reference. 
Within the implicit solvation
framework\cite{cramerImplicitSolvationModelsCR1999}, hydration free energy
of a molecule is often approximated\cite{honigMacroscopicModelsAqueousJPC1993} as the sum of polar (electrostatic) and non-polar components:

\begin{align}\label{eq:thermo_decomposition}
\Delta G_{solv} &= \Delta G_{el} + \Delta G_{np}.
\end{align}
Of the two components in equation \ref{eq:thermo_decomposition}, the
\textit{electrostatic solvation free energy}, $\Delta G_{el}$, often 
contributes the most to the total in polar solvents such as water, especially for macromolecules.
Highly approximate, yet computationally efficient, ways to estimate the non-polar component, $\Delta G_{np}$, are widely used; by comparison, $\Delta G_{el}$ 
is relatively expensive to estimate computationally with commonly used numerical methods such as NPB\cite{onufrievWaterModelsBiomolecularWIRCMS2017}.
Here, our focus is $\Delta G_{el}$.
For a discrete charge density indexed by $i$ we can compute $\Delta G_{el}$ as \cite{onufrievWaterModelsBiomolecularWIRCMS2017}:
\begin{align}\label{eq:electrostatic_solvation_energy}
\begin{aligned}
\Delta G_{el} &= \frac{1}{2} \sum\limits_i q_i \left[ V( \mathbf{r}_i ) - V( \mathbf{r}_i )_{vac} \right],
\end{aligned}
\end{align}
where $V(\mathbf{r}_i)$ and $V(\mathbf{r}_i)_{vac}$ are the 
electrostatic potentials due to the given charge
distribution in the solvent and in vacuum, respectively,
sampled at each point charge $q_i$ located at $\mathbf{r}_i$.
In the special case when the inner and outer dielectric constants, $\epsilon_{in}$ and $\epsilon_{out}$, are equal to $1$ and $80$, respectively, we call $\Delta G_{el}$ the \textit{electrostatic hydration free energy}.
We use equations \ref{eq:standard_potential_two_dielectric} and \ref{eq:electrostatic_solvation_energy} to write:

\begin{align}\label{eq:asc_electrostatic_solvation_energy}
\Delta G_{el} &= \frac{1}{2} \sum\limits_i q_i \left( \oint_{\partial S} \frac{ \sigma( \mathbf{s} ) }{ \vert \mathbf{r}_i - \mathbf{s} \vert }\, d^2s \right).
\end{align}
Though equation \ref{eq:asc_electrostatic_solvation_energy}
is valid for any choice of DB, the surface integral is
non-trivial to compute.
We approximate the surface integral using a specific
triangulation of the DB, see section \ref{ssec:DB}.
This discrete representation approximates equation \ref{eq:asc_electrostatic_solvation_energy} as:

\begin{align}\label{eq:discrete_electrostatic_solvation_energy}
 \Delta G_{el} &\approx \left[ \sum\limits_i \sum\limits_T \frac{ q_i \sigma_T A_T }{ \vert \mathbf{r}_i - \mathbf{r}_T \vert } \right],
\end{align}
where $\sigma_T, A_T$, and $\mathbf{r}_T$ are the apparent
surface charge, area, and center of the triangle $T$ (to express $\Delta G_{el}$ in kcal/mol, which is often convenient, the equation above is multiplied by
166, while using atomic units of length, \AA, and charge, $|e|$ ).
The ASC on $T$ is found by averaging the ASC at its comprising vertices.
The triangular center is simply the centroid of $T$, with its area calculated using Heron's formula:
\begin{align}\label{eq:triangle_area}
  A = \sqrt{ d \cdot ( d - a ) \cdot ( d - b ) \cdot ( d - c ) } \quad ; \quad d = \frac{ a \cdot b \cdot c }{2},
\end{align}
where $a,b,$ and $c$ are the side lengths of $T$.

\section{Numerical Applications and Results}\label{sec:numerical_applications_results}

\subsection{Analytical ASC Computational Speed}\label{ssec:computational_speed}

Here we present general running time descriptions for each tested method, rather 
than exact time values.
In this way, we can differentiate between each method,
without worrying about particular optimizations and expert
parameter set-ups that can be found across a variety of implementations\cite{aguilarEfficientComputationTotalJCTC2012}.
\begin{table}[H]
	\begin{center}
		\resizebox{\linewidth}{!}{
		\begin{tabular}{c|c|c|c}
\textbf{Method} & \textbf{Small Molecules (Average of
16 atoms)} & \textbf{2LZT
(1958 atoms)} & \textbf{DNA (1598 atoms)}\\\hline\hline
\textbf{IGB5(AMBER)} & milliseconds & $\sim$ a second & $\sim$ half a second\\\hline
\textbf{Analytical ASC Approximation} & $\sim$ 100
milliseconds &  tens of seconds & $\sim$ half a minute\\\hline 
\textbf{Numerical PB} &  tens of seconds &  minutes &  tens of minutes\\
		\end{tabular}
		}
	\end{center}
	\caption{
Running time expectations for computed electrostatic solvation free energies.
Times are given per-molecule (averaged over the entire set in the small molecule case).
	}
	\label{tab:wall_time_result_table}
\end{table}
In algorithmic time complexity, the three methods we compare in Table \ref{tab:wall_time_result_table} are very different.
GB methods, such as the IGB5 reference, scale quadratically in the number of atoms ($K^2$), while our method grows linearly ($KN$) in the number of atoms and surface elements ($N$).
Volumetric methods, similar to the NPB reference, scale cubically in the number of grid points per side of a corresponding bounding box, itself a function of grid density and the maximum intra-molecular distance.
The impact of these asymptotic time complexities can be clearly seen when
we focus on hen-egg lysozyme (2LZT) and double-stranded DNA wall running times.
Though the 2LZT structure has about 400 more atoms than the DNA structure, the intra-molecular width of the DNA structure is almost double that of 2LZT.
This means that the DNA structure has both a larger total
surface area and requires a bigger volumetric bounding box;
as seen in Table \ref{tab:wall_time_result_table}, we find longer running times for our ASC approximation and the NPB reference for 2LZT as compared to DNA, but not for the IGB5 reference.
Hence, this contrasting algorithmic complexity affects computational timings between each model for structures of different size and characteristic.
A principal consequence is the following:
The best case scenario for the efficiency of our ASC approximation is for structures having many atoms, but a comparatively low surface area; in terms of the derivation of our model, it is coincidental that in three dimensions the shape maximizing total inner ``volume'' (number of atoms) and minimizing outer surface area is that of a sphere.

Though the efficiency of our analytical ASC implementation is not at the level of the IGB5 reference, it occupies a different niche: its main purpose is the estimation of the ASC and the electric field. 
It is worth noting that the surface integration required for equation \ref{eq:discrete_electrostatic_solvation_energy} is trivially parallel, since equations \ref{eq:electric_field_normal} and \ref{eq:sigma_total} can be computed independent of adjacent surface elements on the DB.
This is in addition to the more fundamental parallelism present in equations \ref{eq:electric_field_normal} and \ref{eq:sigma_total}, due to the independence of per-charge contributions to the electric field and ASC, respectively.
Improvements in the efficiency of our implementation would greatly
improve performance as surface resolution is increased, or in
``worst-case'' scenarios such as those seen in the DNA fragment, Table \ref{tab:wall_time_result_table}.


\subsection{Analytical ASC Accuracy with respect to the NPB Reference}\label{ssec:AASC_NPB_test}

\subsubsection{Small Molecules}\label{sssec:sm_test}

\paragraph{Qualitative Visualizations}\label{par:sm_qualitative}

We first examine how our ASC approximation compares to the NPB reference
\textit{qualitatively}.
Here, our motivation is the well recognized utility\cite{tolokhWhyDoublestrandedRNANAR2014,honigClassicalElectrostaticsBiologyS1995,wangCrystalStructureSARSCoV22020} of visualizing electrostatic characteristics of molecules, including mapping them onto a molecular surface. 
Our main metric, in this section, is qualitative similarity, or lack thereof, between visualizations produced by our analytical ASC method and the NPB reference.

\begin{figure}[H]
	\begin{center}
		\resizebox{\linewidth}{!}{
		\begin{subfigure}{0.33\textwidth}
			\subcaption{}
			\includegraphics[width=\textwidth]{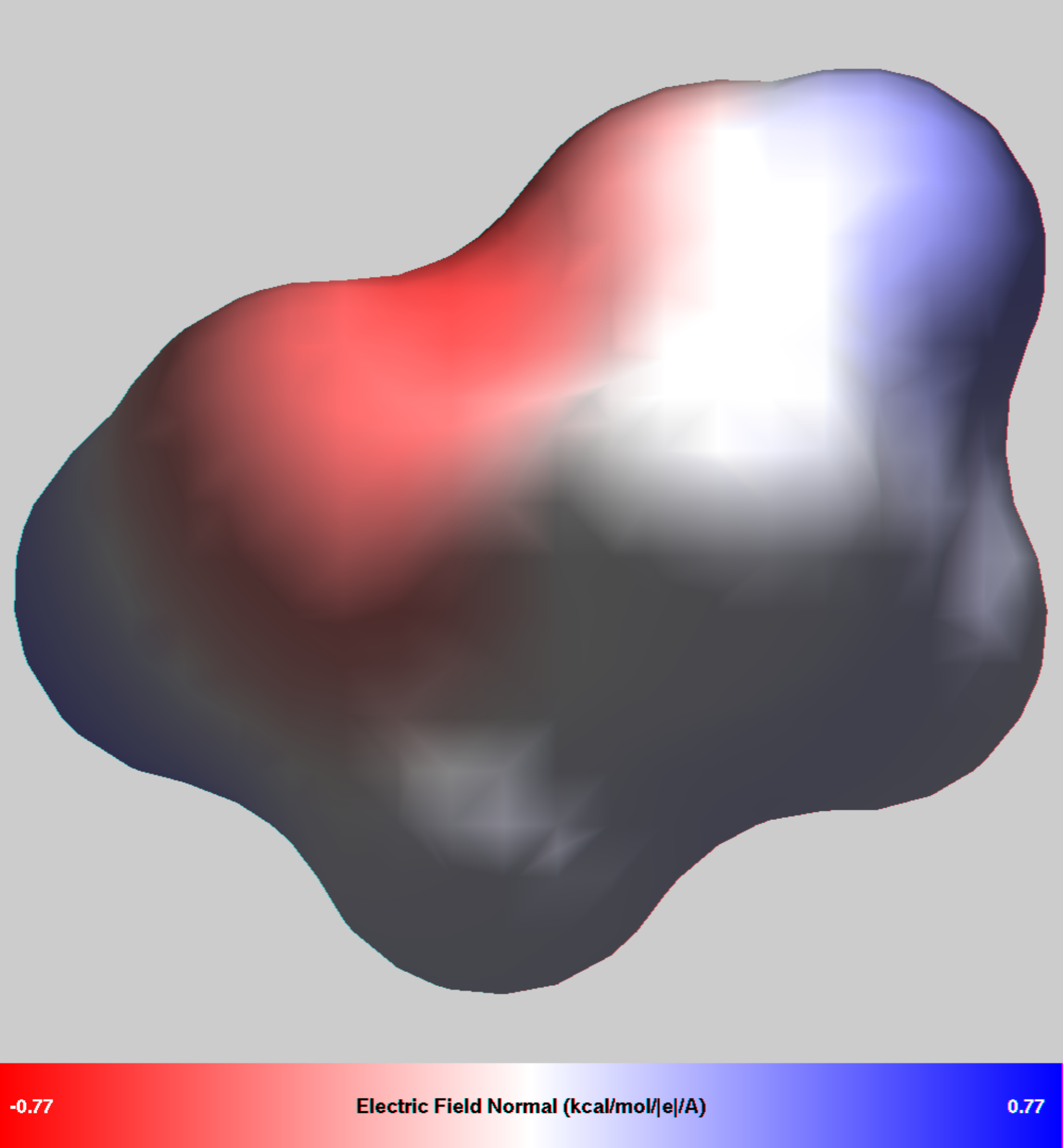}
			\label{sfig:12_ethanediol_GEM}
		\end{subfigure}
		\begin{subfigure}{0.33\textwidth}
			\subcaption{}
			\includegraphics[width=\textwidth]{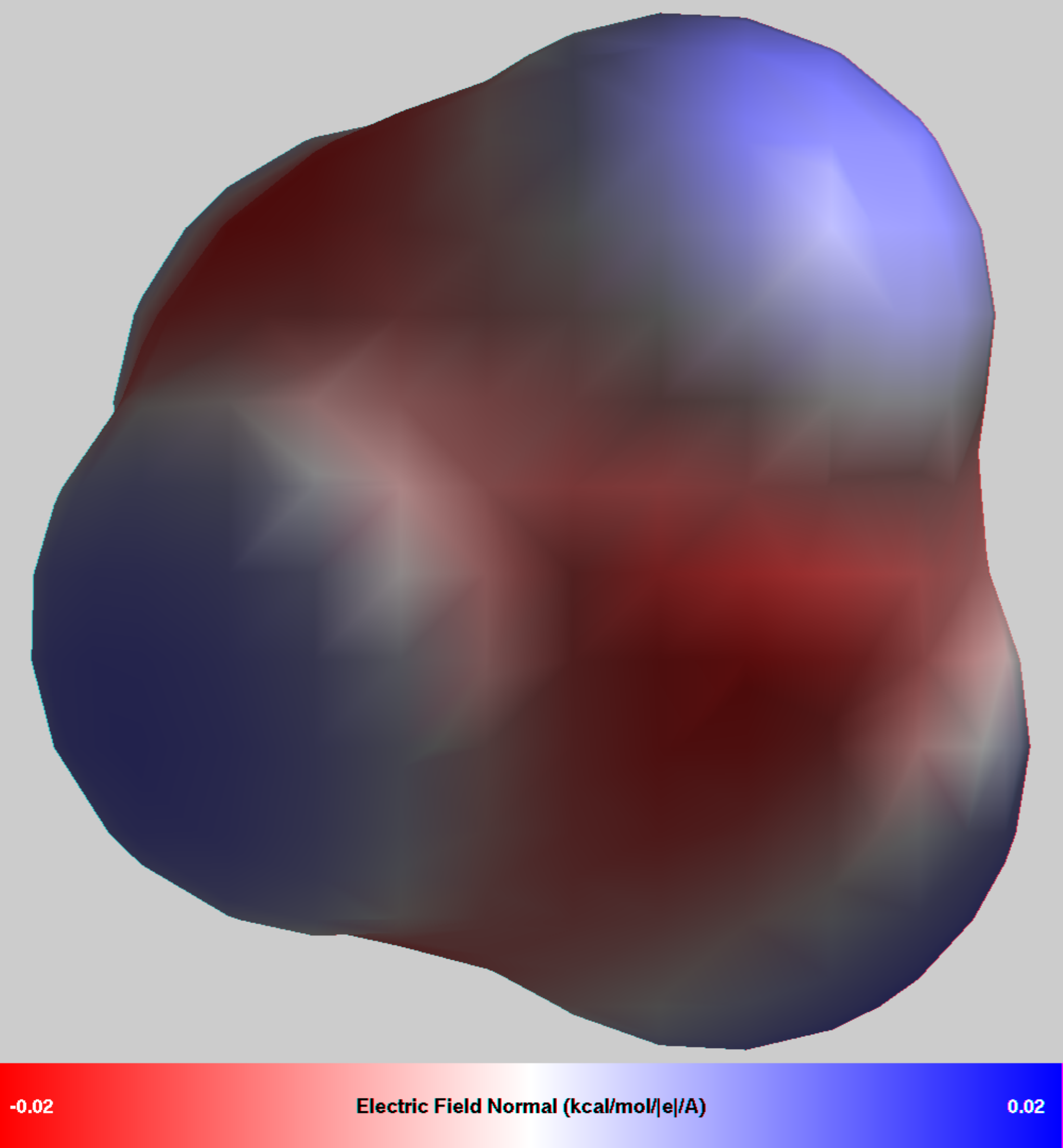}
			\label{sfig:methane_GEM}
		\end{subfigure}
		\begin{subfigure}{0.33\textwidth}
			\subcaption{}
			\includegraphics[width=\textwidth]{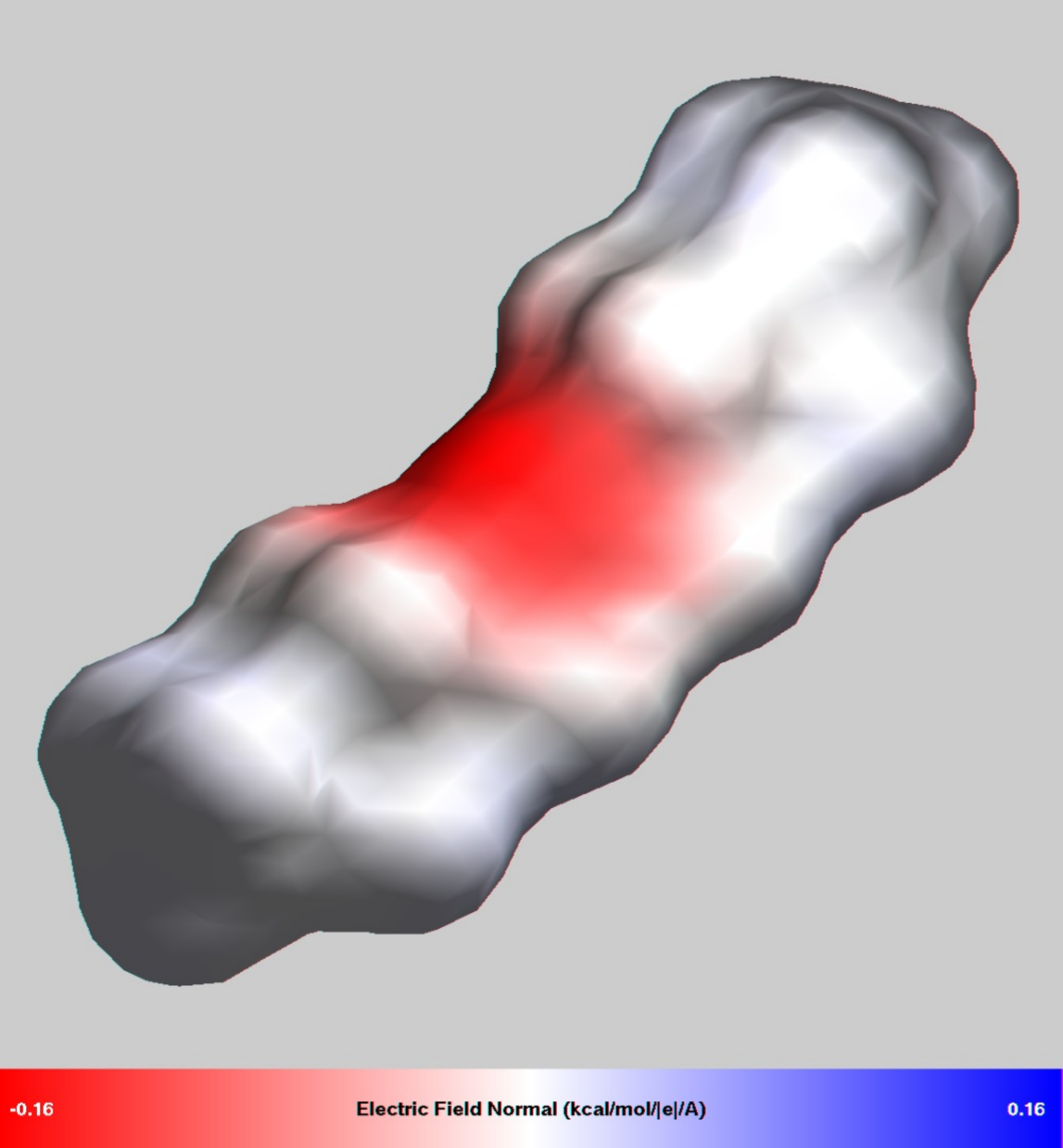}
			\label{sfig:di_n_butyl_ether_GEM}
		\end{subfigure}
		}
		\resizebox{\linewidth}{!}{
		\begin{subfigure}{0.33\textwidth}
			\subcaption{}
			\includegraphics[width=\textwidth]{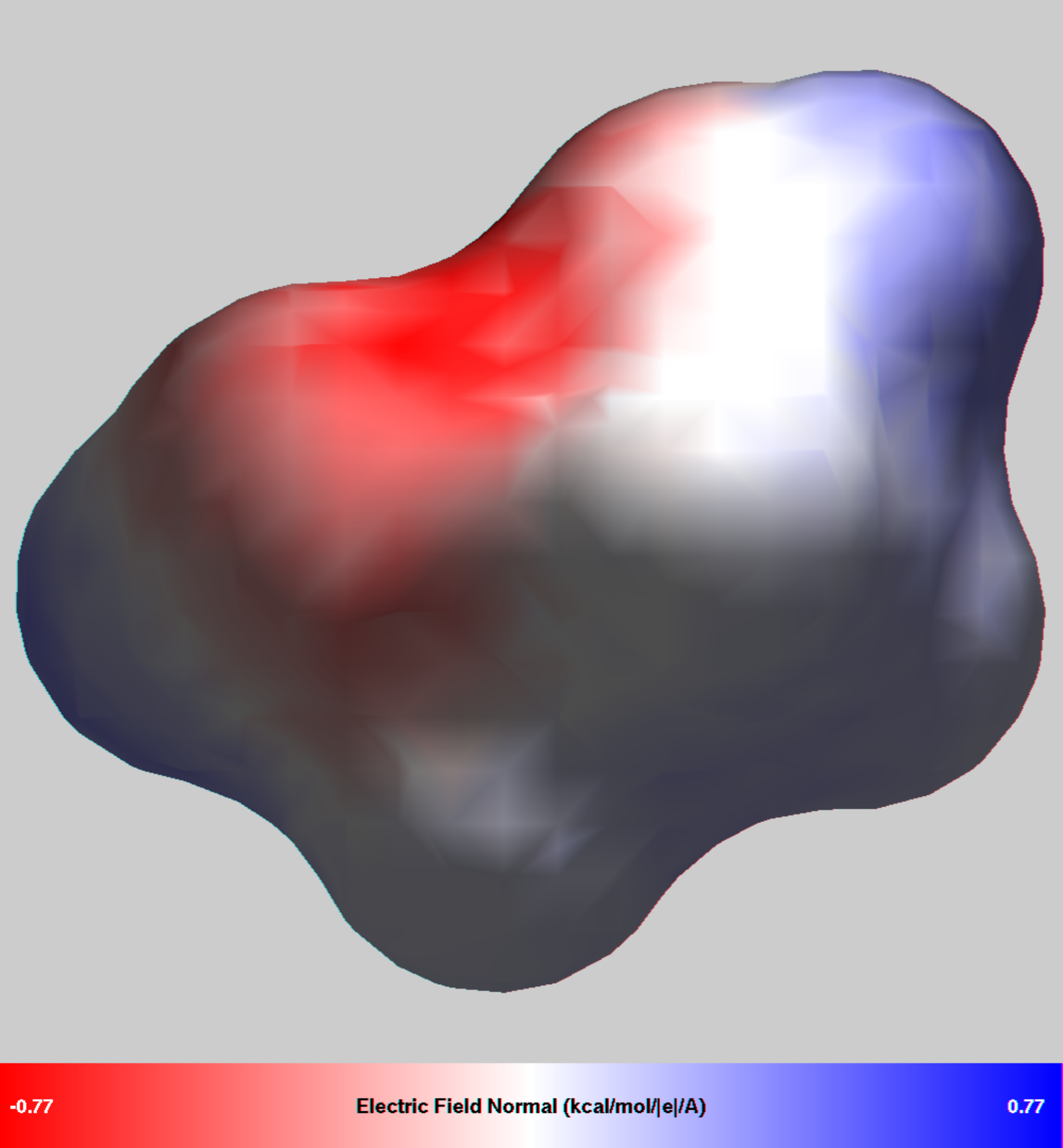}
			\label{sfig:12_ethanediol_MEAD}
		\end{subfigure}
		\begin{subfigure}{0.33\textwidth}
			\subcaption{}
			\includegraphics[width=\textwidth]{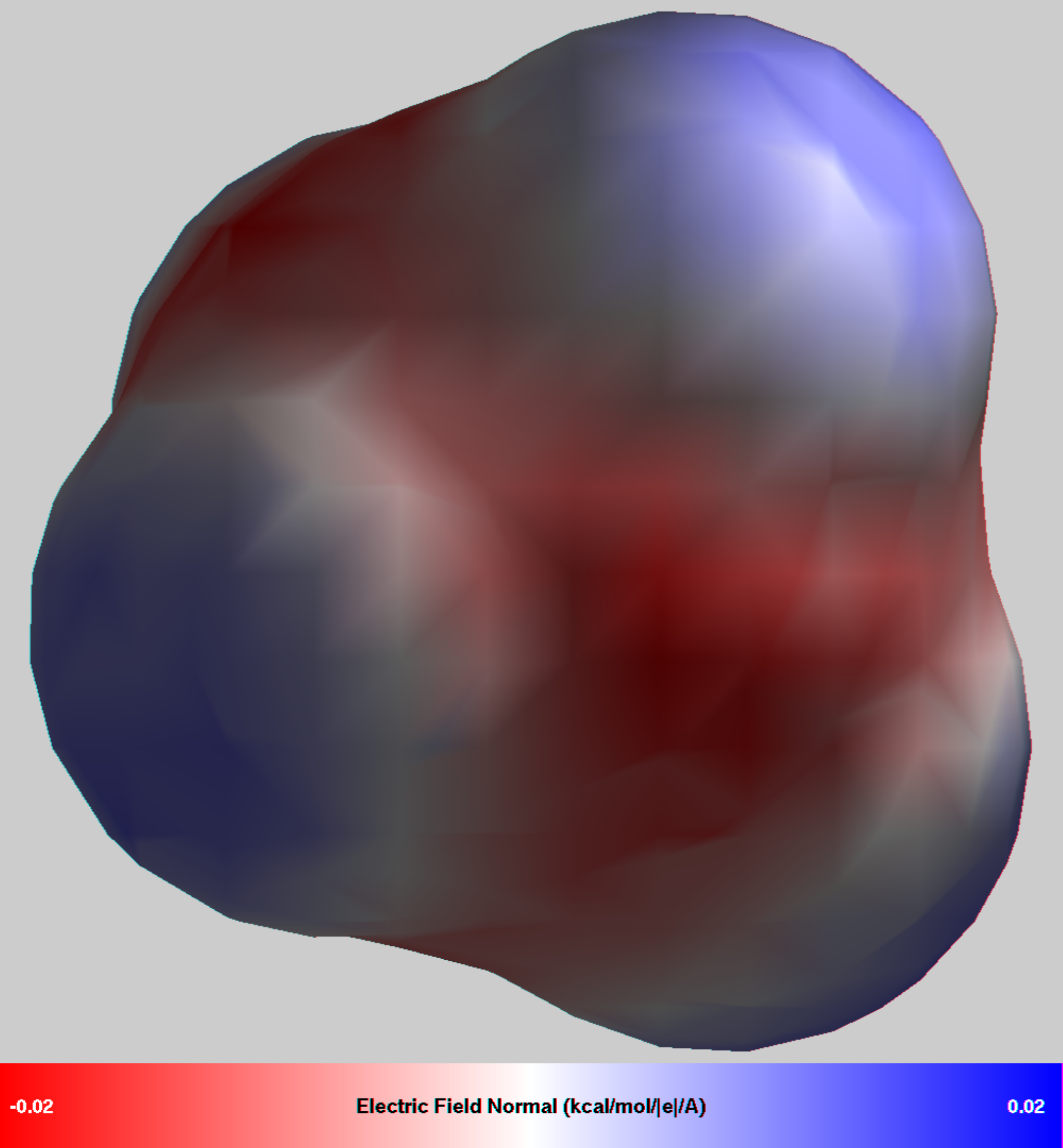}
			\label{sfig:methane_MEAD}
		\end{subfigure}
		\begin{subfigure}{0.33\textwidth}
			\subcaption{}
			\includegraphics[width=\textwidth]{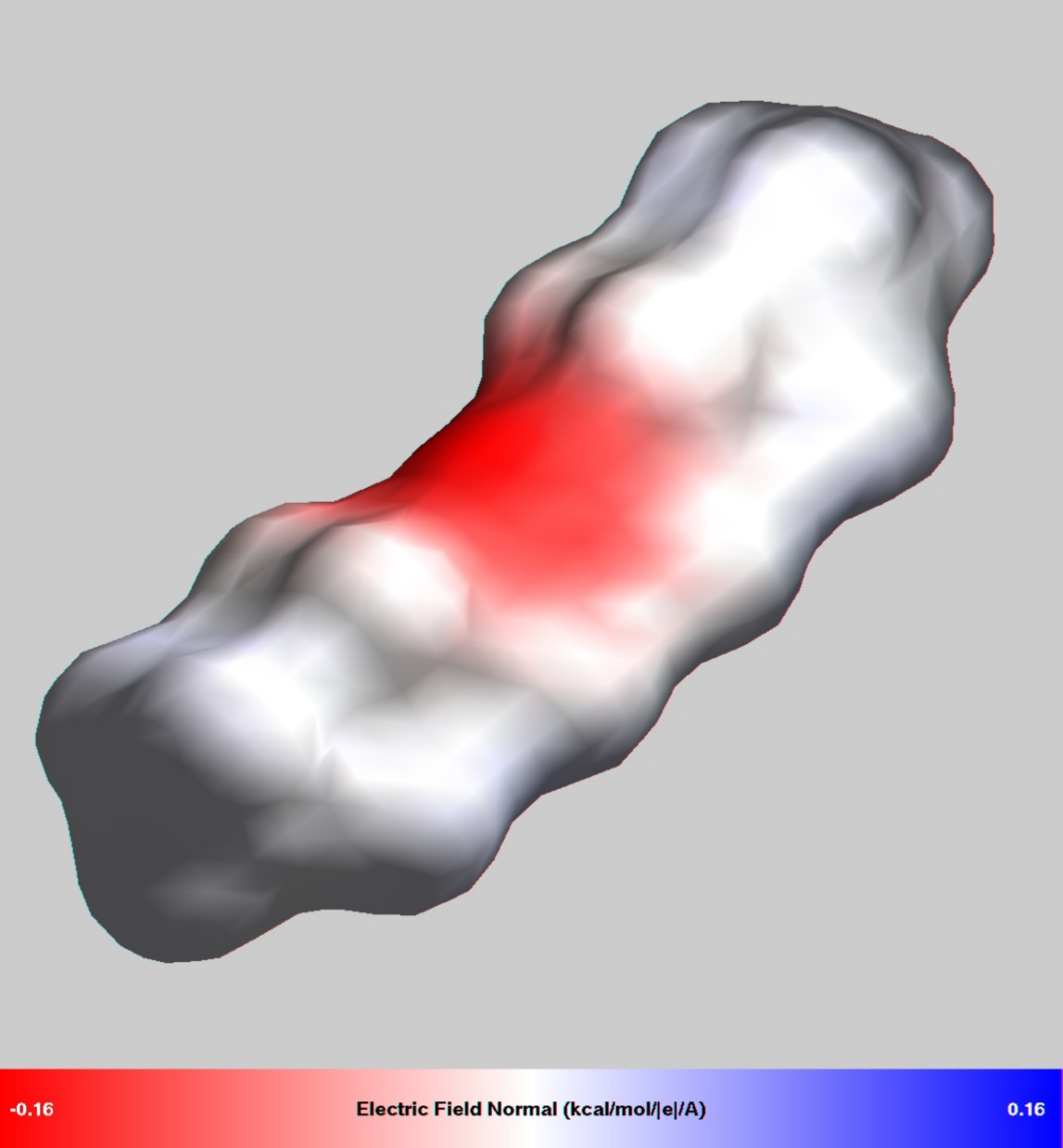}
			\label{sfig:di_n_butyl_ether_MEAD}
		\end{subfigure}
		}
	\end{center}
	\caption
	{
Electric field normals computed on a selection of small molecules by our ASC approximation (top row) and the NPB reference (bottom row), with visualization by GEM\cite{gordonAnalyticalApproachComputingJCP2008}.
From left to right, the three molecules shown (chosen to represent various shapes) are 1,2-ethanediol (\ref{sfig:12_ethanediol_GEM}, \ref{sfig:12_ethanediol_MEAD}), methane (\ref{sfig:methane_GEM}, \ref{sfig:methane_MEAD}), and 1-Butoxybutane (\ref{sfig:di_n_butyl_ether_GEM}, \ref{sfig:di_n_butyl_ether_MEAD}).
All calculations are made $0.7$ \AA~from the DB, with a water probe radius
of $1.4$ \AA. Our ASC approximation and the NPB reference use a $0.1$
\AA~triangulation density/grid spacing. The color range for
the analytical ASC and the NPB reference is the same for each vertical
pair of panels.
	}
	\label{fig:2x3_GEM_MEAD_Comparison_SM}
\end{figure}
Figure \ref{fig:2x3_GEM_MEAD_Comparison_SM} exemplifies the qualitative match between our method and the NPB reference on a subset of the small molecule data set.
Apart from very small irregularities, attributable to the discrete sampling of the NPB potential map, the electric field normals computed with our ASC approximation are, visually, almost indistinguishable from those computed with the NPB reference.

The near quantitative agreement between the analytical ASC and the NPB
reference, Figure \ref{fig:2x3_GEM_MEAD_Comparison_SM} is nontrivial, and
goes beyond the visual match of the ``reds" and the ``blues" with the NPB
reference. 
Notice that, when $\alpha$ is set to $0$ in equations
\ref{eq:fenley_potential} and \ref{eq:electric_field_normal}, the
analytical ASC reduces to the so-called \textit{Coulomb Field
Approximation} (CFA), which is often used, and can be considered a Null
model here.


\begin{figure}[H]
	\begin{center}
		\resizebox{\linewidth}{!}{
		\begin{subfigure}{0.33\textwidth}
			\subcaption{}
			\includegraphics[width=\textwidth]{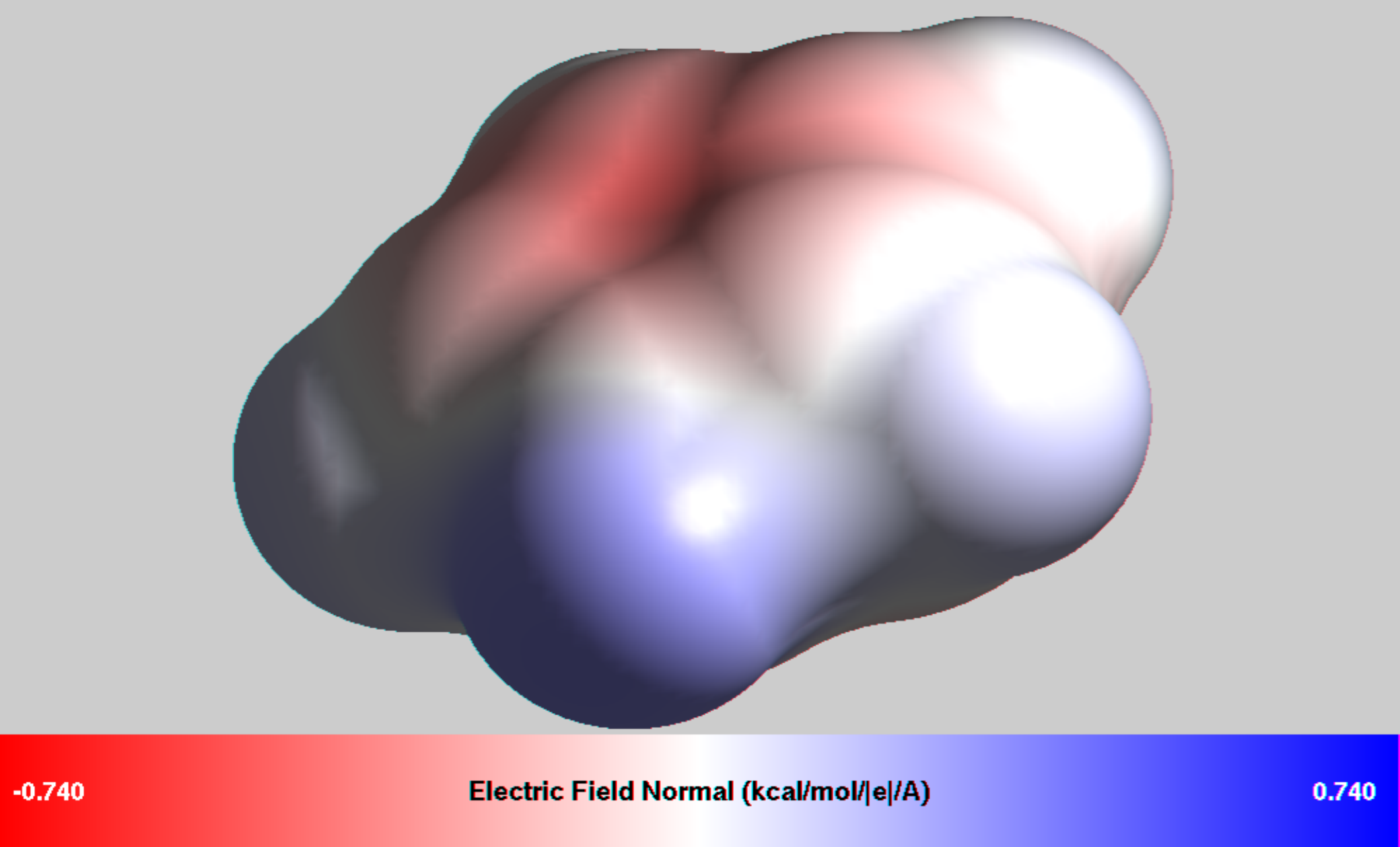}
			\label{sfig:pyrolle_AASC}
		\end{subfigure}
		\begin{subfigure}{0.33\textwidth}
			\subcaption{}
			\includegraphics[width=\textwidth]{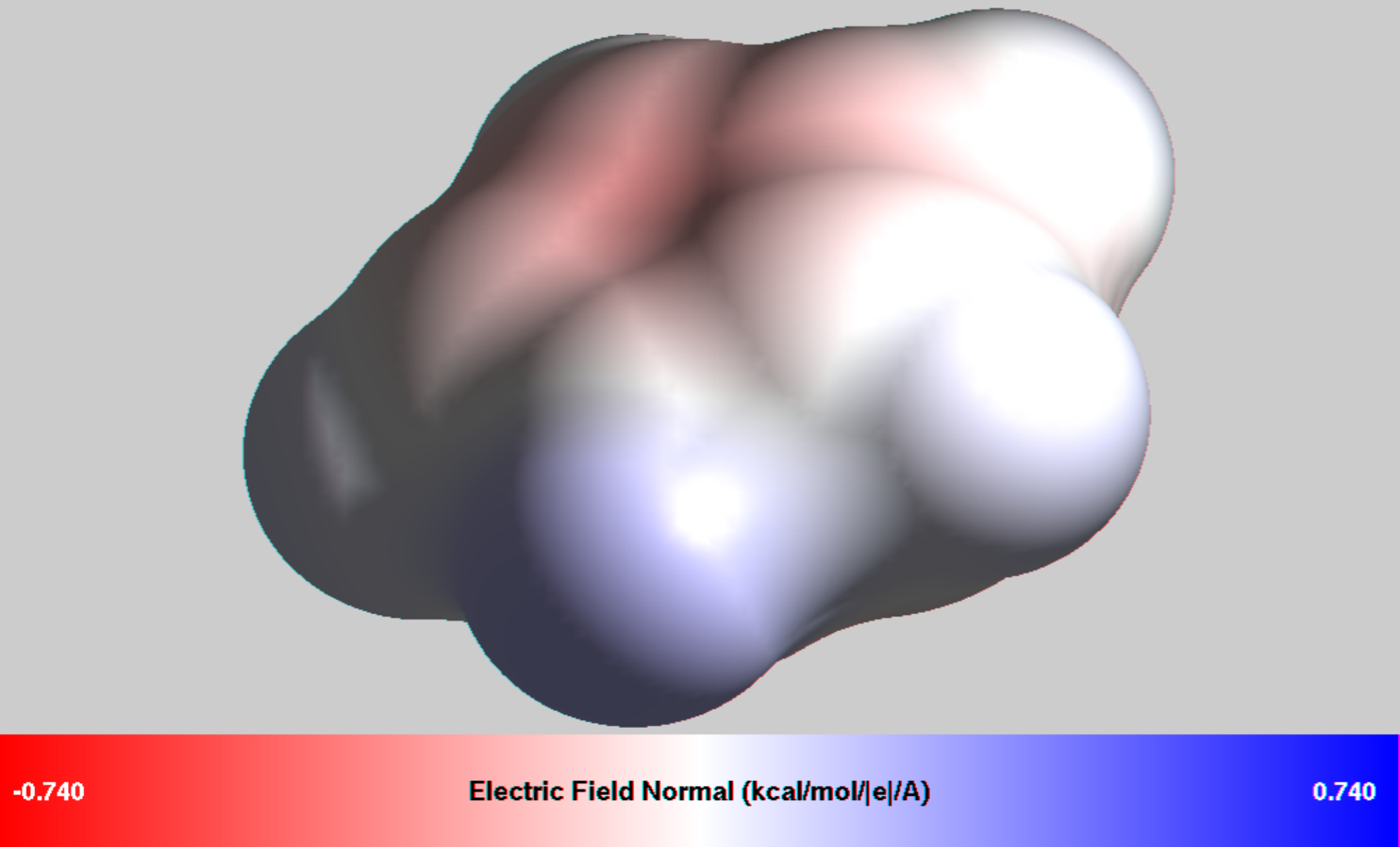}
			\label{sfig:pyrolle_CFA}
		\end{subfigure}
		\begin{subfigure}{0.33\textwidth}
			\subcaption{}
			\includegraphics[width=\textwidth]{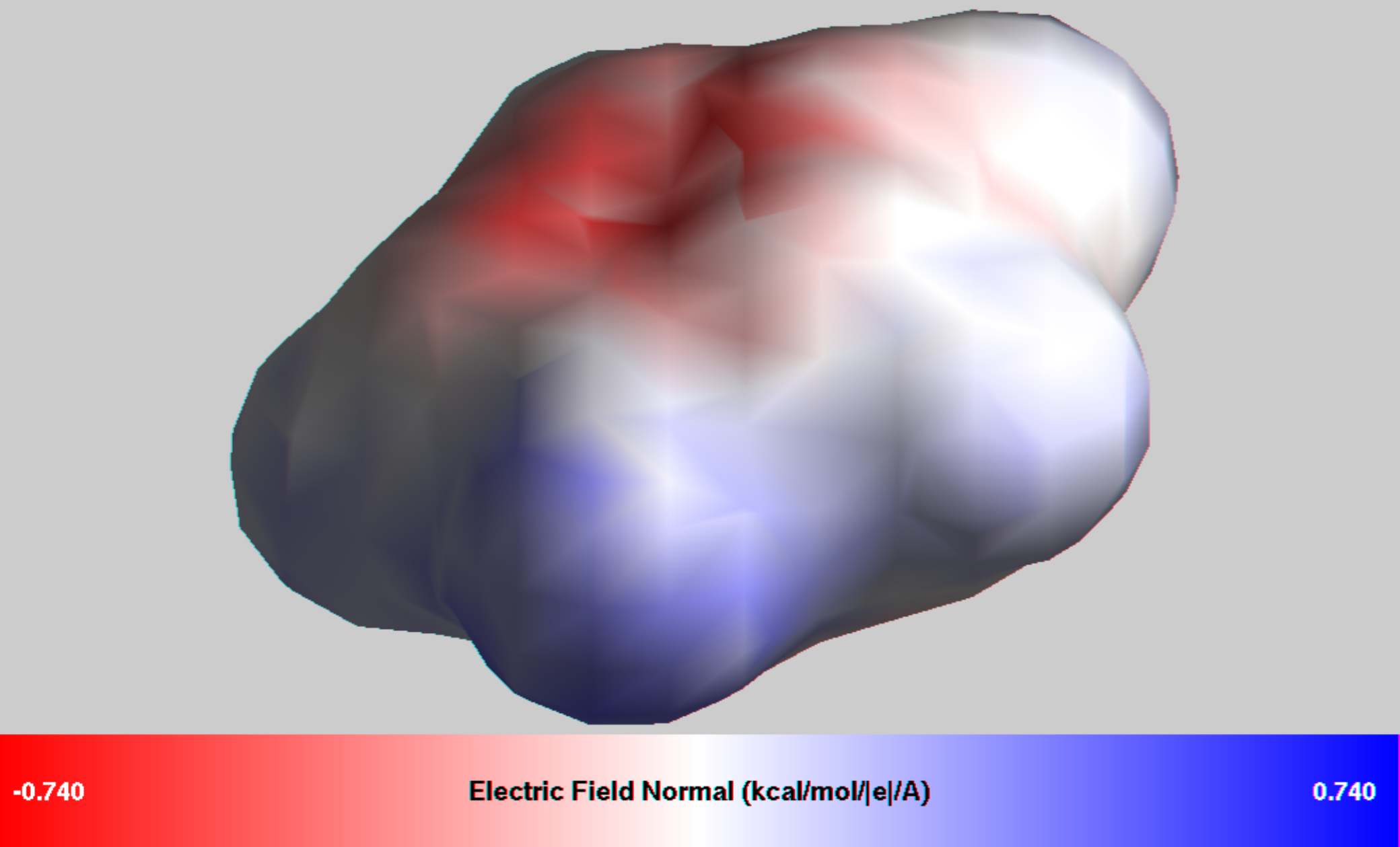}
			\label{sfig:pyrolle_MEAD}
		\end{subfigure}
		}
	\end{center}
	\caption
	{
Electric field normals computed via our Analytical ASC method
(Panel \ref{sfig:pyrolle_AASC}), the \textit{Coulomb Field Approximation}
(CFA) (Panel \ref{sfig:pyrolle_CFA}), and the NPB reference (Panel
\ref{sfig:pyrolle_MEAD}) on a member of our small molecule data set,
pyrrole. While both the analytical ASC and the CFA identify the positive
and negative surface charge patches, the CFA underestimates their intensity
significantly. 
Between the analytical ASC method and the CFA, minimum, maximum, and
average electric field normals on the DB are $-0.441, 0.314$, and $-0.030$
kcal/$( \text{ mol } \cdot e \cdot \text{\AA})$ and $-0.281, 0.200$, and
$-0.019$ kcal/$( \text{ mol } \cdot e \cdot \text{\AA})$, respectively.
The color range used to visualize the field is the same in all the panels. 
	}
	\label{fig:1x2_AASC_CFA}
\end{figure}

Figure \ref{fig:1x2_AASC_CFA} clearly indicates the decreased magnitude of
CFA-generated electric field normals, when compared to our ASC
approximation; on average, the CFA produces an electric field that is
approximately $36$\% weaker than our analytical ASC method, which
reproduces the NPB reference closely. The significant deviation of the CFA
surface charge from the reference translates into its poor accuracy in
estimation of the hydration free energy, Figure \ref{fig:HFE_scatterplot}
below. 

\paragraph{Quantitative assessment of ASC accuracy}\label{par:sm_quantitative}


Next, we examine how our ASC approximation compares
\textit{quantitatively} to the NPB reference.
Though electric field normals (or linearly related surface charges) are 
not the most intuitive accuracy metrics, these quantities are those that 
our method \textit{directly} computes, and so a direct comparison with the 
NPB is in order.
In section \ref{sssec:discussion}, we discuss a physical interpretation of 
the deviation of these quantities from the reference.
Relative to the NPB reference, our ASC approximation achieves an average RMSD and absolute difference of 0.14 and 0.11 kcal/$( \text{ mol } \cdot e \cdot \text{\AA})$ on calculated electric field normal values, respectively.

\paragraph{Electrostatic Solvation Free Energy}\label{par:sm_ESFE}


Relative to direct comparisons of molecular ASC, or its proxy, the normal electric field, tests featuring the calculation of electrostatic solvation free energies are valuable in the sense that they provide an intuitive accuracy metric, directly relevant to experiment. 
Here, IGB5 \cite{onufrievExploringProteinNativePSFB2004} is 
an example of what can be expected from a very fast GB model on small molecule data sets \cite{katkovaAccuracyComparisonSeveralJoMGaM2017,izadiAccuracyComparisonGeneralizedJCTC2018}, in terms of accuracy and running time.
We test how our ASC approximation and the IGB5 method compares to calculation of electrostatic solvation free energies by the NPB reference.

\begin{figure}[H]
		\resizebox{\linewidth}{!}{
      %
      \includegraphics[width=\textwidth]{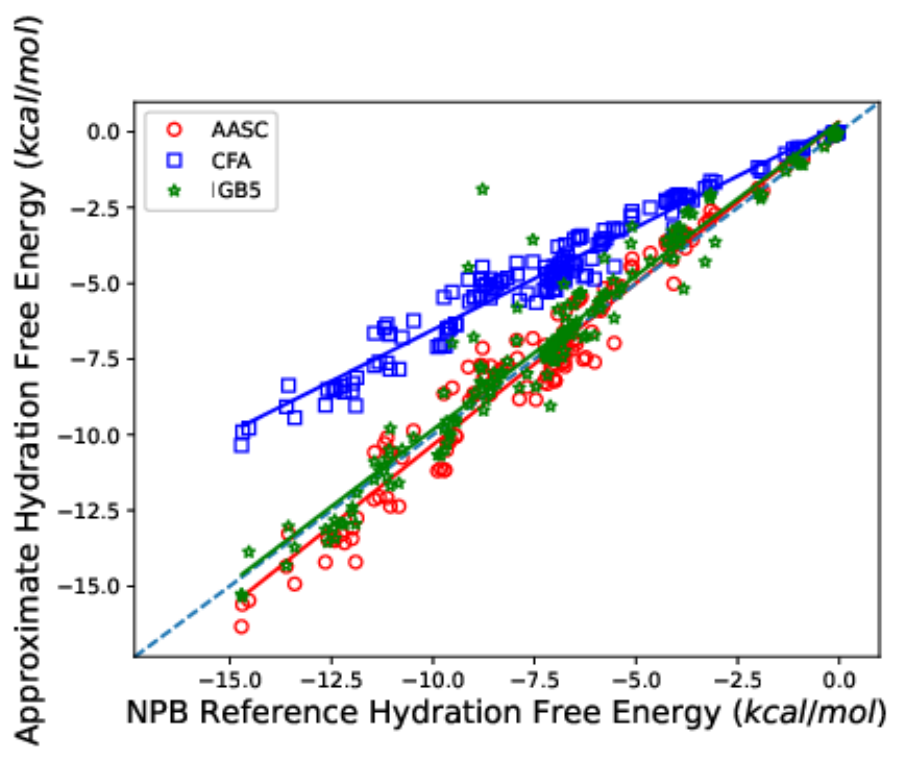}
		}
	\caption{
Analytical ASC (red circles), the GB (green stars) and CFA (blue
squares) hydration free energies of rigid small molecules relative to the
NPB reference values. 
Linear regression lines are plotted using colors corresponding to the representative data points.
The dotted blue line represents the perfect match between an 
approximation and the NPB reference.
$R^2$ values for IGB5, our analytical approximation, and the CFA are $0.932$, $0.961$, and $0.777$, respectively.
Hydration free energies are shown in kcal/mol.
	}
	\label{fig:HFE_scatterplot}
\end{figure}

Using the analytical ASC approximation given in equation \ref{eq:sigma_total}, $\Delta G_{el}$ is computed through the use of equation \ref{eq:asc_electrostatic_solvation_energy}.
Our ASC approximation yielded a 0.77 kcal/mol RMSD to
NPB reference hydration free energies, while the GB reference showed 0.98
kcal/mol on the same metric. Not surprisingly, the CFA accuracy is much
worse, 2.73 kcal/mol away from the NPB reference. 
  In this comparison, NPB reference small molecule electrostatic hydration free energies ranged from -0.02 to -14.71 kcal/mol.

From our findings above, our ASC approximation has an approximately 20\% reduced RMSD to NPB reference hydration free energies than the IGB5 reference.
This accuracy gain is encouraging, particularly from the perspective of design: GB models are designed to analytically approximate the electrostatic solvation free energy obtained from solving the Poisson equation. 
That our model can estimate solvation energies more
accurately than a widely used GB model suggests that the
approximation of the ASC and electric field normals by the model is reasonably
accurate to be considered for practical applications. 
Additionally, the results achieved above give another encouraging conclusion, with respect to running times efficiencies.
To achieve a deviation from the NPB reference of just slightly above $kT$, 
we do not require an overly fine triangulation density for the analytical ASC.
When the grid resolution is set to what is used in the timing
section above (Table \ref{tab:wall_time_result_table}), RMSD against the NPB reference, differs only in non-significant digits.
Thus, our analytical ASC  can achieve a very similar accuracy, without incurring a heavy 1-2 order of magnitude time penalty, as seen with the NPB reference at this fine grid resolution.

\subsubsection{Proteins and DNA}\label{sssec:lm_test}

In analyzing the performance of our model on structures of
increased size and complexity, we examine a fragment of 
double-stranded DNA and the hen-egg lysozyme.
Structures of this type - with regions of the DB having deep, negative curvature ``pockets" - present some of the toughest tests for our model, due to certain theoretical considerations we touch on below.

\paragraph{Qualitative Visualizations}\label{par:lm_qualitative}

\subparagraph{Double-Stranded DNA.}\label{par:DSDNA}

First, we examine our analytical approximation on a double-stranded DNA
fragment.
\begin{figure}[H]
	\begin{center}
		\resizebox{\linewidth}{!}{
		\begin{subfigure}{0.49\textwidth}
			\subcaption{}
			\includegraphics[width=\textwidth]{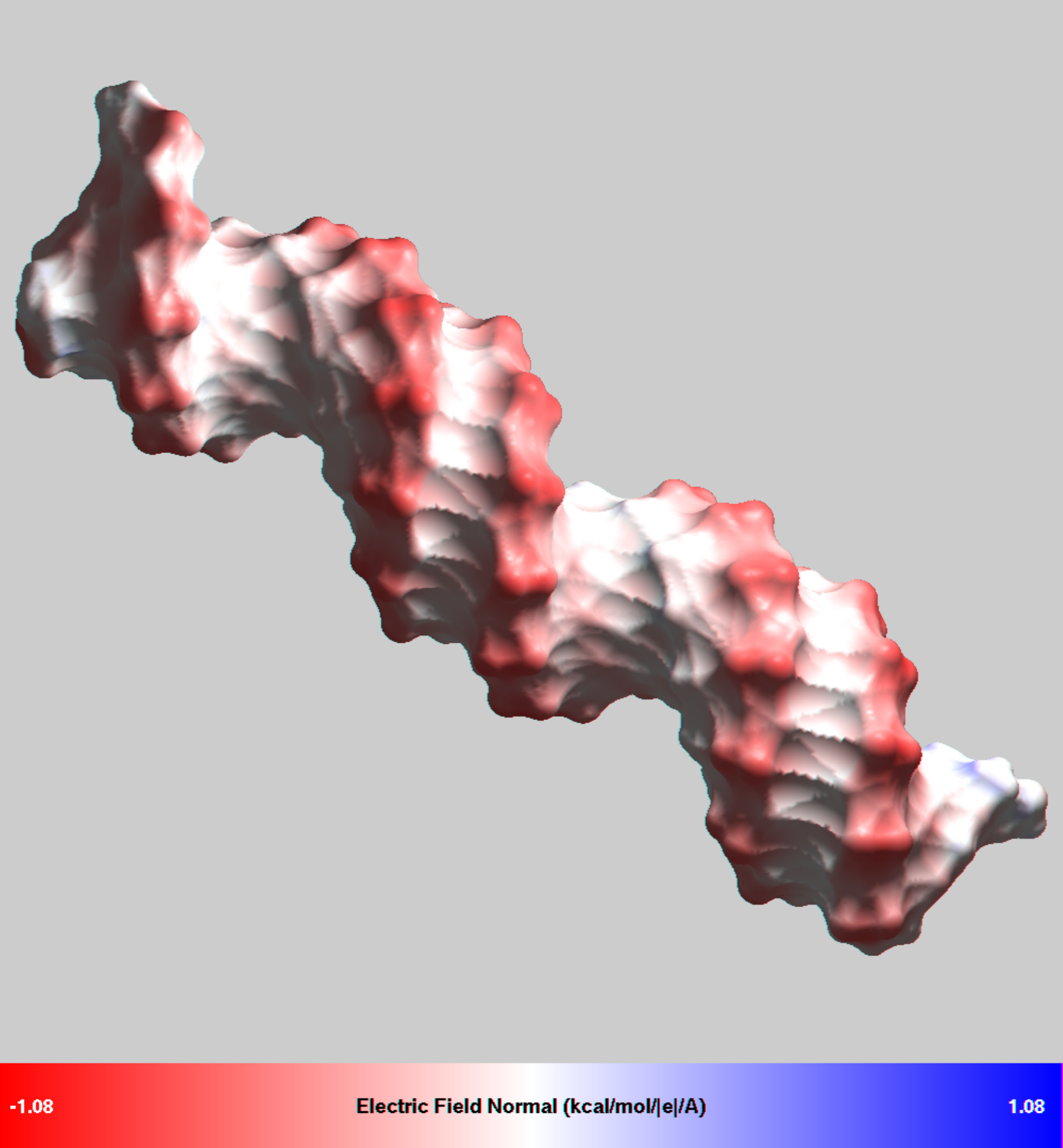}
			\label{sfig:DNA_GEM}
		\end{subfigure}
		\begin{subfigure}{0.49\textwidth}
			\subcaption{}
			\includegraphics[width=\textwidth]{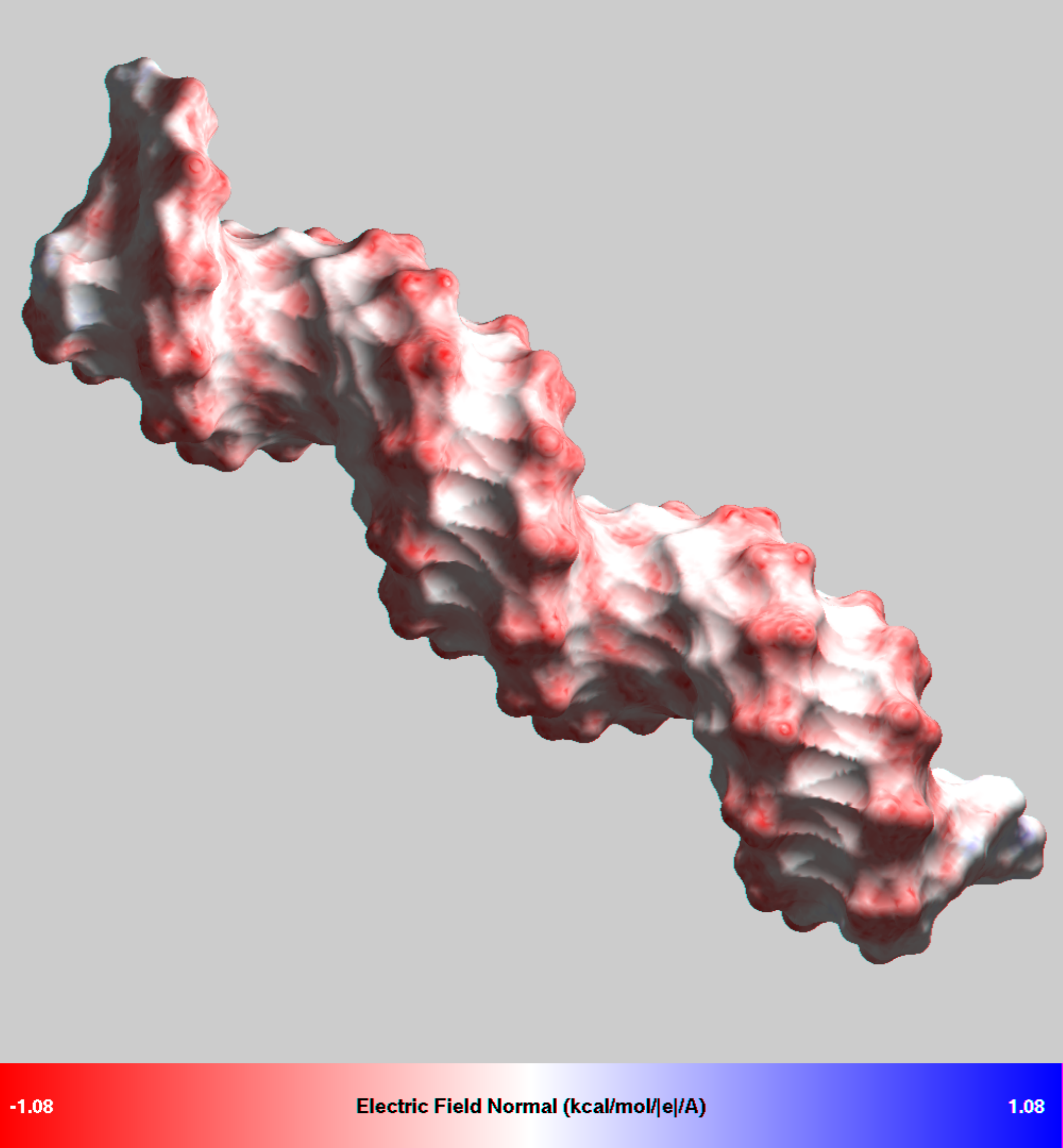}
			\label{sfig:DNA_MEAD}
		\end{subfigure}
		}
	\end{center}
	\caption
	{
Electric field normals computed on the double-stranded DNA snapshot by our ASC approximation, Panel \ref{sfig:DNA_GEM}, and the NPB reference, Panel \ref{sfig:DNA_MEAD}, with visualization by GEM\cite{gordonAnalyticalApproachComputingJCP2008}.
The field is estimated  $1.5$ \AA~from the DB, obtained
with the water probe of radius of $2$ \AA.
The larger probe radius is used here to achieve a better visualization of negative curvature regions.
Our ASC approximation and the NPB reference use a $0.5$ \AA~triangulation density/grid spacing.
	}
	\label{fig:2x2_GEM_MEAD_Comparison_DNA_poly}
\end{figure}
Qualitatively, Figure
\ref{fig:2x2_GEM_MEAD_Comparison_DNA_poly} shows that our
analytical approximation reproduces the NPB reference quite
well, although some discrepancies are clearly seen in the grooves, that is
in the regions of negative curvature, see a discussion below.  

\subparagraph{Triclinic Hen Egg White Lysozyme}\label{par:HEL}

Next, we compare our analytical ASC  to the NPB reference on the triclinic hen egg white lysozyme.
\begin{figure}[H]
	\begin{center}
		\resizebox{0.9\linewidth}{!}{
		\begin{subfigure}{0.49\textwidth}
			\subcaption{}
			\includegraphics[width=\textwidth]{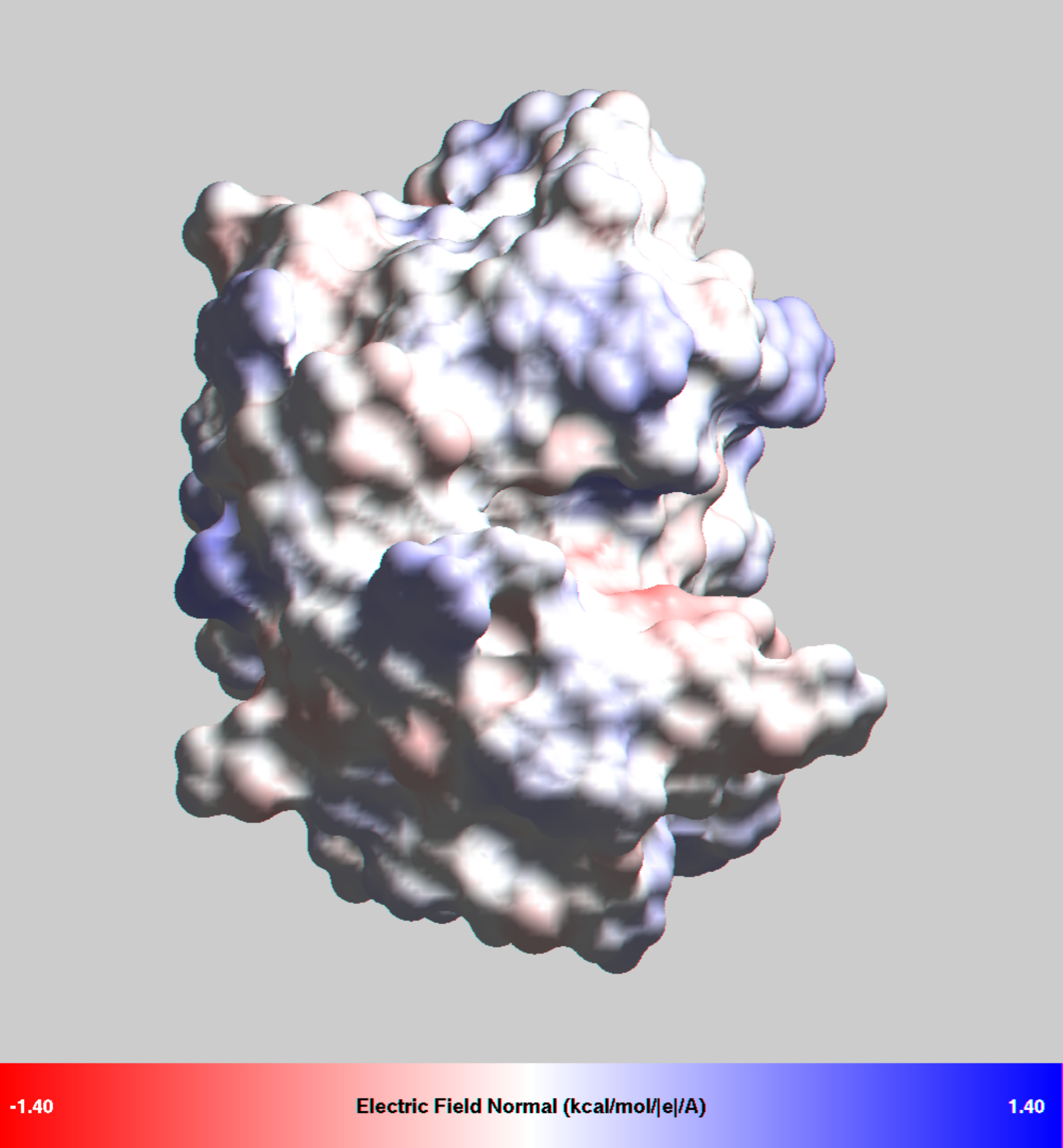}
			\label{sfig:2LZT4_5_GEM}
		\end{subfigure}
		\begin{subfigure}{0.49\textwidth}
			\subcaption{}
			\includegraphics[width=\textwidth]{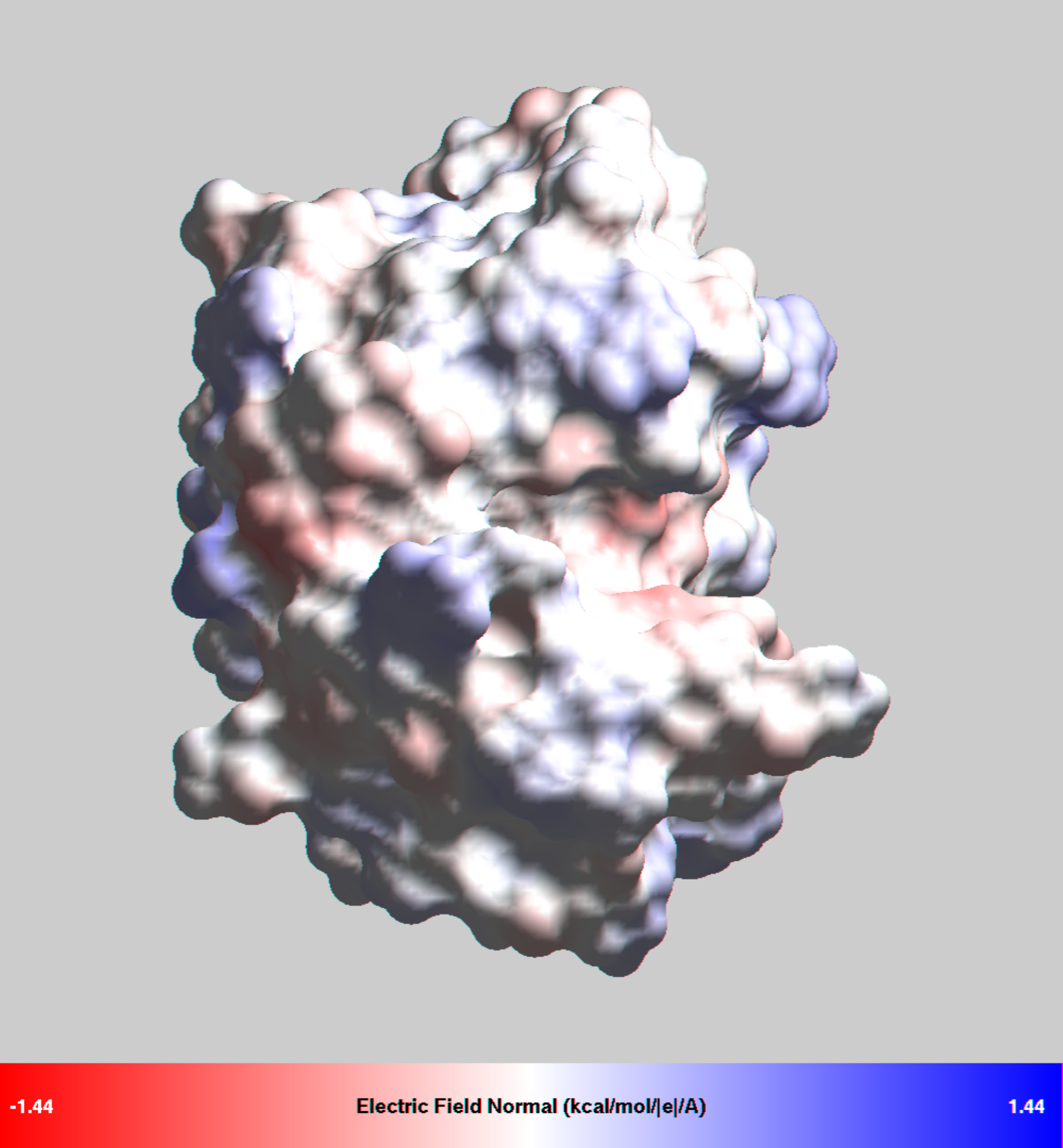}
			\label{sfig:2LZT6_5_GEM}
		\end{subfigure}
		}
		\resizebox{0.9\linewidth}{!}{
		\begin{subfigure}{0.49\textwidth}
			\subcaption{}
			\includegraphics[width=\textwidth]{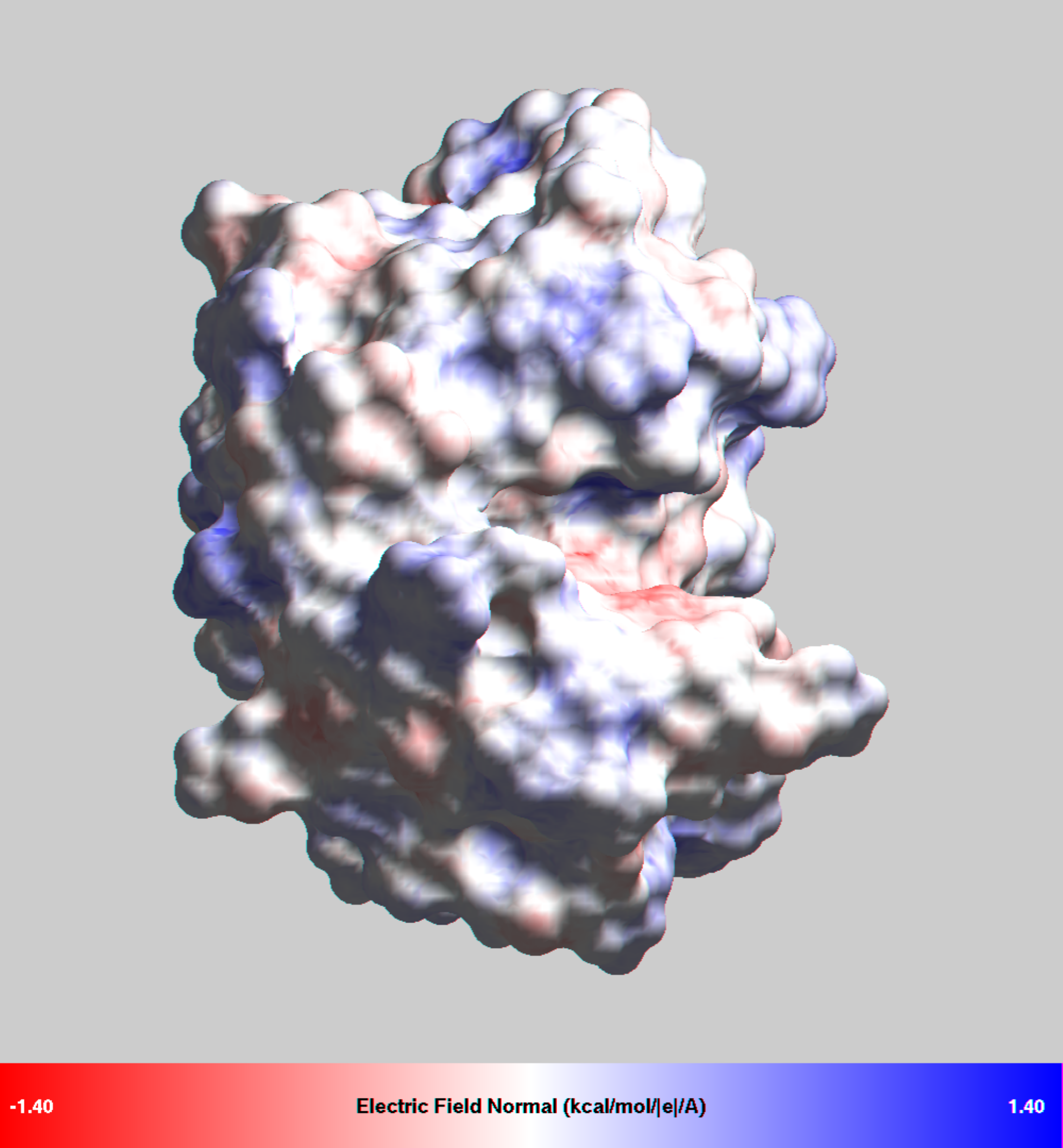}
			\label{sfig:2LZT4_5_MEAD}
		\end{subfigure}
		\begin{subfigure}{0.49\textwidth}
			\subcaption{}
			\includegraphics[width=\textwidth]{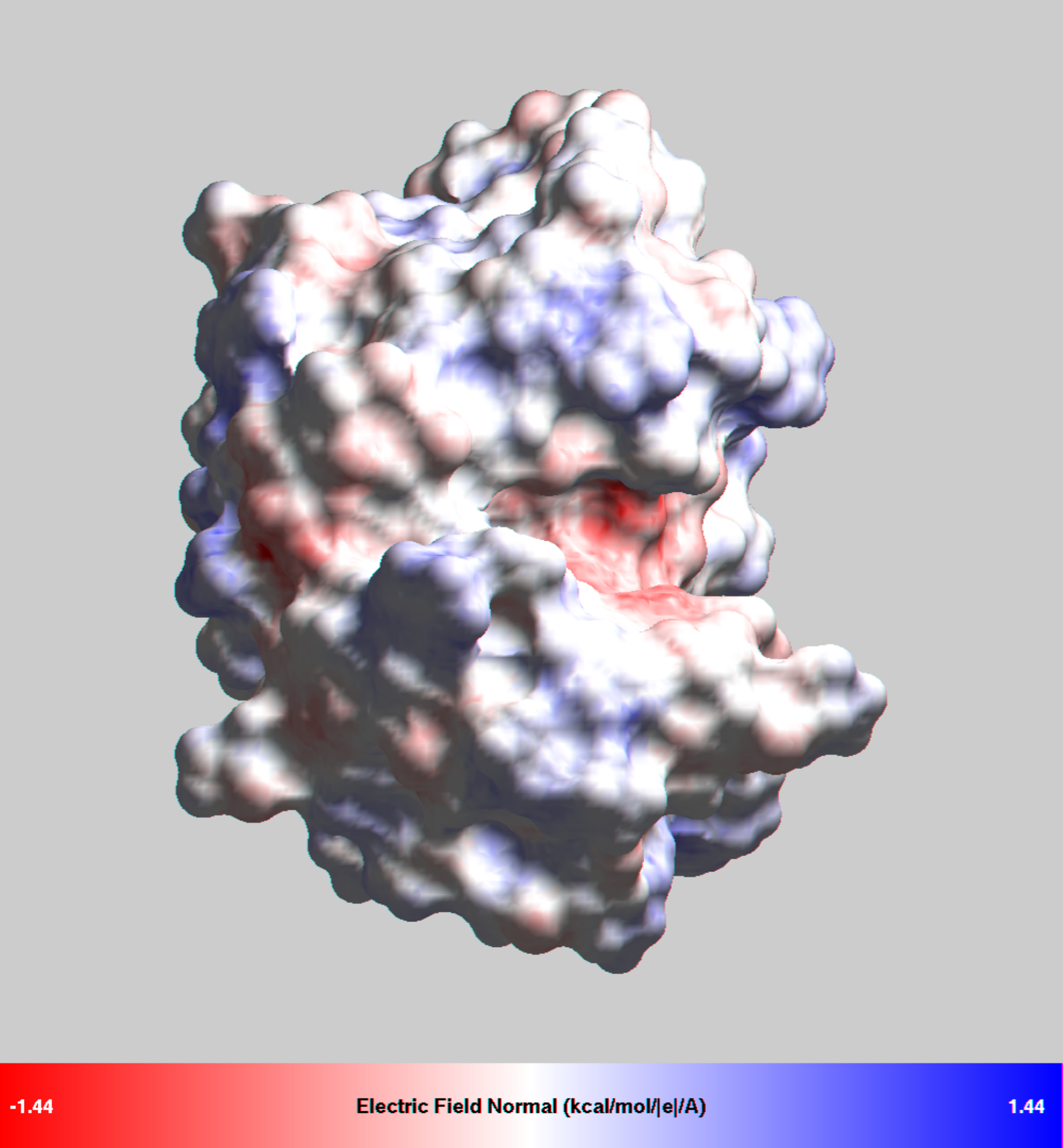}
			\label{sfig:2LZT6_5_MEAD}
		\end{subfigure}
		}
	\end{center}
	\caption
	{
Electric field normals computed on the hen-egg lysozyme by our ASC approximation (top row) and the NPB reference (bottom row), with visualization by GEM\cite{gordonAnalyticalApproachComputingJCP2008}.
Panels \ref{sfig:2LZT4_5_GEM} and \ref{sfig:2LZT4_5_MEAD}: the structure at pH 4.5.
Panels \ref{sfig:2LZT6_5_GEM} and \ref{sfig:2LZT6_5_MEAD}: the structure at pH 6.5.
All calculations are made 1.5 \AA~from the DB, with a water probe radius of 2 \AA. 
The larger probe radius is used to achieve a better visualization of negative curvature regions.
Our ASC approximation and the NPB reference use a $0.5$ \AA~triangulation density/grid spacing.
	}
	\label{fig:2x2_GEM_MEAD_Comparison_2LZT}
\end{figure}
We see in Figure \ref{fig:2x2_GEM_MEAD_Comparison_2LZT} that our analytical approximation accurately reproduces the NPB reference, outside of the hen-egg lysozyme's binding cleft.
Within the binding cleft, quantitative deviations in
electric field normal magnitudes from the NPB reference
become apparent, though our approximation still produces
a qualitatively reasonable picture.
Our approximation (Figure \ref{sfig:2LZT4_5_GEM} $\rightarrow$
\ref{sfig:2LZT6_5_GEM}) qualitatively reproduces the reference (Figure \ref{sfig:2LZT4_5_MEAD}
$\rightarrow$ \ref{sfig:2LZT6_5_MEAD}), in visualizing the substantial electrostatic
effect of Asp 52 and Glu 35 in the enzymatic pocket, as well as the
corresponding changes
due to the change in the charge states of these two residues 
under mildly acidic conditions (pH
4.5).
Hence, both the NPB reference and our model are able to
visualize the the behavior of Asp 52 and Glu 35, under pH change\cite{blakeCrystallographicStudiesActivityPotRSoLSBBS1967,phillipsThreedimensionalStructureEnzymeSA1966,warshelTheoreticalStudiesEnzymicJoMB1976}.

\paragraph{Quantitative Comparisons}\label{par:lm_quantitative}

With qualitative tests complete, we finish the analysis with a quantitative comparison between our analytical approximation and the NPB reference.
\begin{table}[H]
	\begin{center}
		\resizebox{\linewidth}{!}{
		\begin{tabular}{c||c|c|c}
& \textbf{Double-Stranded DNA} & \textbf{2LZT pH 4.5} & \textbf{2LZT pH 6.5}\\\hline\hline
\textbf{Absolute Difference} & 0.27 & 0.15 & 0.15\\\hline
\textbf{Average RMSD} & 0.37 & 0.19 & 0.20\\
		\end{tabular}
		}
	\end{center}
	\caption
	{
Electric field normal comparisons between our analytical approximation and the NPB reference, on double-stranded DNA and the protonated/un-protonated hen-egg lysozyme.
All values are in kcal/$( \text{ mol } \cdot e \cdot \text{\AA})$.
	}
	\label{tab:ef_large_result}
\end{table}
As expected, quantitative performance deficiencies exist for larger molecules with prominent regions of negative curvature.
Although the double-stranded DNA and hen-egg lysozyme have
similar numbers of atoms, average RMSD values in Table \ref{tab:ef_large_result} are, relatively, inconsistent.
On the double-stranded DNA snapshot ($\sim$1600 atoms), the
average RMSD against the NPB reference is about 2.6 times larger than on small molecules (section \ref{par:sm_quantitative}, Table \ref{tab:ef_large_result}).
Comparatively, the average RMSD of the hen-egg lysozyme ($\sim$2000 atoms) is only about 1.4 times larger than on small molecules.
On hydration free energies, relative errors between our analytical approximation and the NPB reference are quite small on the DNA snapshot, $\sim$ 4\%, but more than double, $\sim$ 12\%, on the hen-egg lysozyme.
These findings might have to do with the DNA's proportion of negative curvature regions with respect to the whole.

\subsubsection{Discussion}\label{sssec:discussion}

\paragraph{Surface Charge Distribution and Electrostatic Solvation Free Energy}\label{par:solvation_context}

To put the results of section \ref{par:sm_quantitative} and Table \ref{tab:ef_large_result} in a better context, it may be helpful to consider a hypothetical situation.
Suppose a biomolecule of interest has a constant electric
field strength near its dielectric boundary.
When compared to the reference, there is, on average,
a $\sim0.4$ kcal/$( \text{ mol } \cdot e \cdot \text{\AA})$ RMSD error in the electric field normal values.
If a unit electric charge was moved $1$ \AA~away from the
biomolecular boundary, along the surface normal, the error
in the total work done by the electric field would be less
than $\sim 0.4$ kcal / mol - small when compared to the ``gold-standard'' 1
kcal / mol difference against reference. Several caveats are due here.
First, the estimate is based on the RMS error; it is possible that the
errors are larger in some spots near the DB. That concern is mitigated to
an extent by the fact that the electric field strength is inversely
related to the square of distance from the surface, and
would, in reality, decrease away from the surface, making
any discrepancy with the reference smaller. Second, this test is free from the
danger of a specific error cancellation that may affect the commonly used
error assessment based on solvation free energies. As the history of the GB
model development demonstrates\cite{Onufriev2000}, very large errors in individual contributions may
cancel out to produce seemingly accurate total solvation energies.

\paragraph{Structural Considerations}\label{par:structural_considerations}

A notable feature, seen prominently in both the
double-stranded DNA snapshot and the hen-egg lysozyme, but
not generally in small molecules, is the presence of
distinct, and fairly deep, negative curvature pockets on the DB.
Our analytical ASC model is derived from an exact solution of the Poisson
problem on a spherical DB (Figure \ref{sfig:one_charge_description}), having positive curvature throughout.
Negative curvature regions, such as the main groove in Figure
\ref{fig:2x2_GEM_MEAD_Comparison_DNA_poly} and the binding cleft in Figure
\ref{fig:2x2_GEM_MEAD_Comparison_2LZT}, do not occur on a sphere; this is
where our model is not expected to perform well. 
A resulting loss in performance had been noted previously
\cite{gordonAnalyticalApproachComputingJCP2008} for the
approximate electrostatic potential (equation \ref{eq:fenley_potential}).
It is therefore surprising that a qualitative agreement
with the NPB reference is still seen, even in these regions.  
More shallow negative curvature regions are also present in the small molecules, Figure
\ref{fig:2x3_GEM_MEAD_Comparison_SM}, but apparently these do not present a
significant challenge to the new model.

Confounding this effect, deep  negative curvature regions
on the DB can restrict water molecule conformational
freedom, making nearby solvent behave less similar to that
of the bulk \cite{liWaterBiomolecularBindingPCCP2007,debeerRoleWaterMoleculesCTiMC2010}.
It has been shown that regions of this type can significantly modify interactions between small molecule inhibitors and their target proteins \cite{spyrakisRolesWaterProteinJMC2017}, prompting investigations related to their identification \cite{hamelbergStandardFreeEnergyJACS2004, petrekCAVERNewToolBB2006}.
In our context, this change in the behavior of water bulk
has negative implications on the performance of our model,
but the same is true for the NPB reference, which is also
based on the continuum solvent.
Because both of these models are expected to deviate from
the correct physical behavior within these regions of
negative curvature, we argue that a qualitative agreement
with the NPB reference may be acceptable here, in place of a strong quantitative agreement. 

\paragraph{Computational Considerations}\label{par:computational_considerations}

Though not implemented in this work, one very important consequence of our analytical approach to the ASC confers a key additional theoretical benefit -- the trivially parallel nature of our solution.
Equation \ref{eq:sigma_total} is computed over our \textit{discrete} DB representation, and is totally independent from surrounding surface elements.
Coupled with a similarly parallelizable expression for the computation of electrostatic solvation free energy, equation \ref{eq:discrete_electrostatic_solvation_energy}, our method holds significant potential for the computationally efficient ASC treatment of large biomolecules.

\subsection{Testing on a Large Biomolecule}\label{ssec:6M0J}

For a real-world application, we examine our approximation on a much larger ($\sim$6500 atom) complex, with important relevance today - the ACE2/SARS-CoV-2 complex (PDB ID:6M0J).
\begin{figure}[H]
	\begin{center}
		\resizebox{\linewidth}{!}{
		\begin{subfigure}{\textwidth}
			\subcaption{}
			\includegraphics[width=\textwidth,height=13cm]{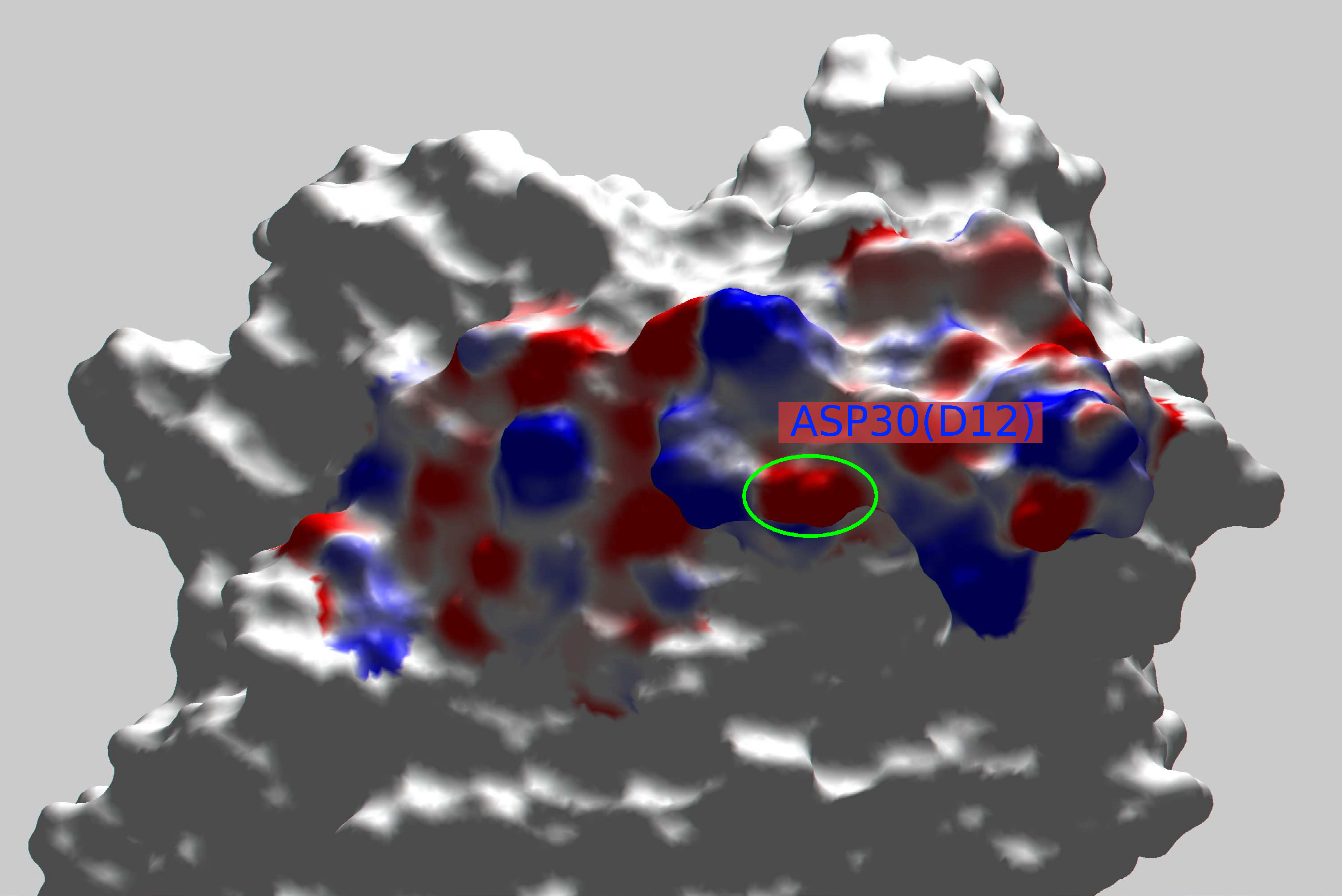}
			\label{sfig:6M0JA_ASC}
		\end{subfigure}
		\begin{subfigure}{\textwidth}
			\subcaption{}
      \includegraphics[width=\textwidth,height=13cm]{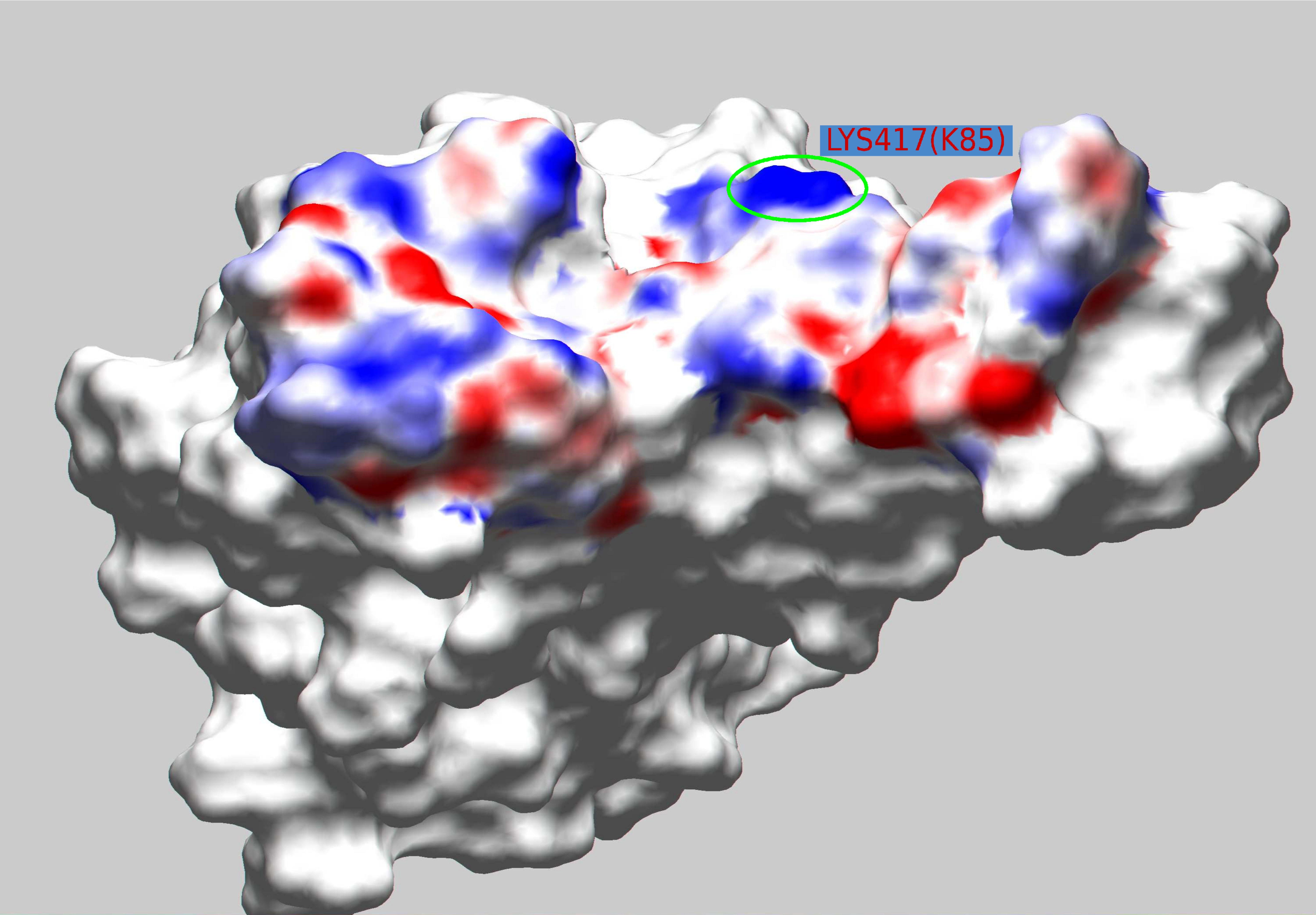}
			\label{sfig:6M0JE_ASC}
		\end{subfigure}
		}
	\end{center}
	\caption
	{
Apparent surface charge computed on the receptor binding domain (RBD) of ACE2 receptor/SARS-CoV-2 spike glycoprotein complex, with visualization by GEM\cite{gordonAnalyticalApproachComputingJCP2008}.
The range of the ASC values is $\pm0.008$ $e/$\AA$^2$, corresponding to the red-blue color map.
White area -- the majority of the molecular surface  outside of the RBD -- is excluded from the calculation, which reduces the computational time significantly.
To mimic the color convention used in Figure 1 (B) of Wang et al\cite{wangCrystalStructureSARSCoV22020}, the sign of equation \ref{eq:sigma_total} is reversed for this calculation, that is the negative of the $\sigma$ is shown.
Figures \ref{sfig:6M0JA_ASC} and \ref{sfig:6M0JE_ASC} show the ACE2 receptor and SARS-CoV-2 spike glycoprotein RBDs, respectively.
All calculations are made $0.7$ \AA~from the DB with a water probe radius of $2$ \AA.
A NanoShaper triangulation density of $0.5$ \AA~is used.
	}
	\label{fig:2x1_6M0J_GEM_ASC}
\end{figure}
Recently, a comparison has been made between SARS-CoV and SARS-CoV-2, examining various mutations and their effects on respective binding strengths with the ACE2 agonist\cite{wangCrystalStructureSARSCoV22020}.
The focus of Figure \ref{fig:2x1_6M0J_GEM_ASC} is on, what
Wang\cite{wangCrystalStructureSARSCoV22020} terms, the ``CR2'' receptor
binding domain; the visualized ASC shows how our approximation reproduces the electrostatic complementary of surfaces charges between two residues, ASP30 (D12), Figure \ref{sfig:6M0JA_ASC}, and LYS 417 (K85), Figure \ref{sfig:6M0JE_ASC}, thought to contribute to the formation of a salt bridge.
This salt bridge improves both stability and binding strength between the ACE2 receptor and SARS-CoV-2 spike protein, when compared to the SARS-CoV spike protein.

These large-scale visualizations of the ASC (or of the
normal component of the electric field) has already been
shown useful\cite{Tolokh2014}; our analytical approximation  might be
a useful tool in understanding complex protein-protein
interactions at a atomistic scale, including SARS-CoV-2 mutants of concern
\cite{zahradnikSARSCoV2RBDVitrob2021}, especially in high throughput
studies.
Our method can potentially be useful in this area due to its targeted,
\textit{source-based} approach to computation of ASC, where only a small
portion the entire DB, and hence a small subset of the surface elements, 
is included in the computation, 
as demonstrated in Figure
\ref{fig:2x1_6M0J_GEM_ASC}. As a result, the compute time is reduced
dramatically, see the discussion on time complexity above. 

\section{Conclusion}\label{sec:conclusion}

In this work, we have derived a closed-form, \textit{analytical}
approximation for biomolecular apparent surface charge (ASC), and the normal
component of the electric field outside the molecule. 
To the best of our knowledge, this is the
first such fully analytical approximation. 
The approximation is constructed to closely reproduce the exact infinite series solution 
for a perfect spherical boundary; more importantly, the approximation yields a reasonably close agreement with the standard numerical PB for realistic molecular structures.
Specifically, quantitative reproduction of results from the standard numerical PB reference 
is achieved on most of the tested molecules, except within prominent
regions of negative curvature, where the new approximation 
is still qualitatively correct.
Comparisons with a popular fast GB model in AMBER (IGB5) shows  
that our method is more accurate in reproducing the hydration free energy,
albeit at higher computational expense, which may be
expected of proof-of-concept code package that is not highly optimized.
At the same time, standard numerical PB is still 1-2 orders of magnitude slower than the proposed
approximation, which puts it ``in-between" fast analytical GB and numerical
PB. We stress that solvation free energy estimates are used here as a
common and convenient accuracy metric, and is not where we believe the
potential benefits of the proposed analytical ASC may be. 
These potential benefits stem from the unique features of the method. 

There are at least two features of the new approximation absent
from the GB: the ability to estimate the apparent surface charge (and,
hence, the potential everywhere); and the ability to estimate the normal
component of the electric field. Another noteworthy feature of the approach sets it
apart from other existing approximations that can estimate ASC, 
including those aimed at computing ASC directly -- the fact that the new approximation is ``source-based".
This means that the normal electric field and the ASC can be estimated at any individual point or surface patch, without the need for self-consistent computation over the entire surface or volume.
This feature is in contrast to ``field-based" methods such as numerical solutions of the Poisson equation or DPCM.
As an illustration, we showed that the ``source-based" feature of our ASC
approximation allows a rapid examination of the ACE2/SARS-CoV-2 RBD
electrostatics, reproducing conditions posited to contribute
to the spike protein's high binding strength. 

	An area which, in our view, can benefit the most from the proposed analytical ASC
is the development of new implicit solvation methods that require fast
estimates of local polarization charges or/and fields.
We also believe the new approach may have the potential to
compete  with existing ASC-based approaches in QM applications,
especially where computational efficiency is key; further
extensive testing and analysis beyond the scope of the proof-of-concept
work will be necessary to explore the potential of the
method in this area.

As it stands, the proposed method has several limitations. First, it does
not yet include salt effects explicitly. However, in the future, it should
be relatively easy to add into the model salt dependence at the Debye-Huckel 
level, following an approach outlined in
Ref. \cite{fenleyAnalyticalApproachComputingJCP2008}. Another limitation of
the model is its qualitative nature in the regions of high negative
curvature, at least relative to the standard NPB reference. Overcoming this
specific limitation will require a significant extension of the underlying
theory, and extensive testing on biomolecular structures.  


A careful and detailed comparison of our proof-of-concept approximation 
within the broader category of existing, 
optimized implementations  ASC
methods has not been performed, and is warranted in the future; 
nonetheless, promising
results so far have pointed to the potential of our approach in forming the basis of novel implicit models of solvation.

\section{Acknowledgments}\label{sec:acknowledgments}

We would like to thank Nitin Passa for his contributions to the numerical ideas behind this application of ASC.
In addition, we appreciate the help of Dr. Igor Tolokh for supplying the double-stranded DNA snapshot, and Dr. Andrew Fenley for his useful comments and suggestions.

\section{Author Declarations}\label{sec:author_declarations}

\noindent A.V.O. reports support from the NIH, grant R21GM131228.

  \subsection{Conflict of Interest}

  \noindent The authors have no conflicts to disclose.

\section{DATA AVAILABILITY} 

The data that support the findings of this study are available from the corresponding author upon reasonable request.

\bibliographystyle{aip}


\end{document}